\documentclass[aps,prl,twocolumn,groupedaddress,showpacs,floatfix,superscriptaddress,longbibliography]{revtex4-2}
\usepackage[plainpages=false,pdfpagelabels,colorlinks=true,linkcolor=red,urlcolor=blue,citecolor=blue,pdftitle={Title},pdfauthor={},pdfdisplaydoctitle=true,pdfduplex=DuplexFlipLongEdge]{hyperref}
\usepackage{siunitx}
\usepackage{mhchem}
\usepackage{epsfig}
\usepackage{graphicx}
\usepackage[utf8]{inputenc}
\usepackage{amsmath,amssymb}
\usepackage[dvipsnames]{xcolor}

\usepackage[normalem]{ulem}

\begin{document}

\title{Breathing-Driven Metal-Insulator Transition in Correlated Kagome Systems}

\author{Qingzhuo Duan}
\affiliation{School of Physics and Astronomy, Beijing Normal University, Beijing 100875, China}
\author{Zixuan Jia}
\affiliation{School of Physics and Astronomy, Beijing Normal University, Beijing 100875, China}
\author{Zenghui Fan}
\affiliation{School of Physics and Astronomy, Beijing Normal University, Beijing 100875, China}
\author{Runyu Ma}
\affiliation{School of Physics and Astronomy, Beijing Normal University, Beijing 100875, China}
\author{Jingyao Meng}
\affiliation{School of Physics and Astronomy, Beijing Normal University, Beijing 100875, China}
\author{Bing Huang}
\affiliation{School of Physics and Astronomy, Beijing Normal University, Beijing 100875, China}
\affiliation{Beijing Computational Science Research Center, Beijing 100084, China}
\author{Tianxing Ma}
\email{txma@bnu.edu.cn}
\affiliation{School of Physics and Astronomy, Beijing Normal University, Beijing 100875, China}
\affiliation{Key Laboratory of Multiscale Spin Physics (Ministry of Education), Beijing Normal University, Beijing 100875, China}

\date{\today}

  \begin{abstract}
    Inspired by the recent discovery of breathing kagome materials \(\rm Nb_3Cl_8\) and \(\rm Nb_3TeCl_7\), we have explored the influence of the breathing effect on the Hubbard model of the kagome lattice. Utilizing the determinant quantum Monte Carlo method, we first investigated the average sign problem in the breathing kagome lattice, which is significantly affected by both the breathing strength and the interaction strength. Secondly, we calculated the electronic kinetic energy, the direct current conductivity, and the electronic density of states at the Fermi level to determine the critical interaction strength for the metal-insulator transition. Our results indicate that the breathing effect, in conjunction with the interaction strength, drives the kagome system from a metal to an insulator. Finally, we evaluated the magnetic properties and constructed a phase diagram incorporating both transport and magnetic properties. The phase diagram reveals that as the interaction strength increases, the system transitions from a paramagnetic metal to a Mott insulator. Our research provides a theoretical guidance for utilizing the breathing effect to control the band gaps, conductivity, and magnetic properties of kagome materials with electronic interactions.
\end{abstract}

\maketitle

\section{I. INTRODUCTION}
\vspace{-0.2cm}

The kagome lattice, distinguished by its unique corner-sharing triangular geometry, stands as a promising platform for investigating numerous correlated electron phenomena, encompassing the metal-insulator transition, magnetism, unconventional superconductivity, and quantum spin liquid states\cite{PhysRevX.11.041010,PhysRevLett.127.236401,PhysRevLett.134.086902,Nature.s41586-021-03983-5,nature08917,Nature.11659,science.aab2120,PhysRevLett.126.247001}.
Over the past two decades, multiple families of kagome materials have garnered significant research attention owing to their fascinating metal-insulator transition behaviors and magnetic properties \cite{NatPhys.s41567-023-02215-z,Nature.s41586-024-07431-y,AdvMater.adma.202410655}. This focus originates from their unique electronic structures and topological features, which render them critical platforms for deciphering the intricate interplay between electron correlations and phase transitions in condensed matter systems \cite{Nature.022.05034,PhysRevX.11.041010}.

Specifically, angle-resolved photoemission spectroscopy (ARPES) was employed to characterize $\rm RbV_3Sb_5$, revealing that its charge density wave phase transition triggers the opening of partial energy gaps and the emergence of new Van Hove singularities \cite{PhysRevX.11.041010,PhysRevLett.127.236401}. Interestingly, the kagome material $\rm Nb_3Cl_8$ was successfully exfoliated via mechanical methods \cite{acs.nanolett.2c00778}, where lattice distortion induced a breathing effect that opened the Dirac cone and introduced multiple band gaps. Subsequently, $\rm Nb_3TeCl_7$ was synthesized using a solid-state approach \cite{adma.202301790}, with theoretical studies establishing that the on-site energy tunability of Nb sites modulates the bandwidth and topological structure of flat bands in this system.  Additionally, the metal-insulator transition can be induced by various modulation techniques. Adjusting the chemical potential in the $\rm (Bi,Sb)_2Te_3$ channels results in a band insulator \cite{PhysRevB.92.155312}. Furthermore, structural distortion under pressure triggers the metal-insulator transition in $\rm CrSiTe_3$ \cite{npj.023.00389}.

In terms of magnetic properties, most kagome materials exhibit antiferromagnetic (AFM) order, such as $\rm ZnCu_3(OH)_6Cl_2$, $\rm FeGe$, and $\rm YMn_6Sn_6$ \cite{PhysRevLett.98.107204,Nature.022.05034,sciadv.abe2680,PhysRevB.103.014416}. Significantly, substituting rare earth atoms in zig-zag chains enables tuning of magnetic properties—for example, transitioning from the nonmagnetic state of $\rm YbTi_3Bi_4$ to the ferromagnetic (FM) phase of $\rm NdTi_3Bi_4$ \cite{CommunMater.024.00513,Nature.022.05516}. FM order has also been observed via ARPES in the $d$-electron kagome metal $\rm Fe_3Sn_2$ \cite{nature25987}. Given the above experimental studies, these advancements highlight the urgent need to elucidate the underlying physical mechanisms governing the metal-insulator transition and magnetism in kagome systems, which has emerged as a pivotal research direction \cite{PhysRevLett.127.177001,PhysRevLett.127.046401}.

A multitude of theoretical approaches have been employed to study the kagome systems. Within the framework of the tight-binding approximation\cite{NatureRevPhys.10.1038}, the hexagonal rotational symmetry of the kagome lattice imparts opposite phases to the eigenfunctions at adjacent corners. This phase opposition leads to a cancellation of the phase during electron hopping between neighboring lattice sites, effectively confining electrons within the hexagonal units. 
The resulting electron localization  gives rise to a pronounced electron correlation effect, and various numerical methodologies have been utilized to investigate the kagome Hubbard model. These include the variational cluster approximation \cite{PhysRevB.93.245123,PhysRevB.98.205114}, cellular dynamical mean-field theory \cite{PhysRevLett.97.066401,PhysRevResearch.2.033476}, variational Monte Carlo studies \cite{CondensedMatter.14.145252,PhysRevLett.133.096501}, and determinant quantum Monte Carlo (DQMC) method \cite{PhysRevB.107.035134,CommunPhys.10.1038}. Collectively, these studies concur that the system experiences a Mott transition at a critical interaction strength. However, the critical interaction strength $U/t$ derived from these investigations range from 5 to 11, a discrepancy that may stem from the diverse strategies employed to refine the ground state approximations.  Beyond kagome lattices, similar mechanisms of correlation-driven localization have been observed in other systems: strong correlations can drive the Lieb and graphene lattices from a metallic state to a Mott insulating state with an energy gap \cite{PhysRevB.108.235163,PhysRevLett.120.116601}, while disorder in lattice-site hopping processes leads to electron localization near the Fermi level, giving rise to an Anderson insulator \cite{PhysRevLett.102.136806,PhysRevB.104.045138}. These findings highlight the universality of electron correlation effects in modulating phase transitions across different lattice geometries.

Despite the diverse modulation techniques for the metal-insulator transition discussed earlier, most kagome materials identified to date are gapless metals. This characteristic inherently prevents them from achieving an ``off" state in devices \cite{Nature.021.01451,Nature.022.05034}, imposing limitations on their applications in logic and optoelectronic systems. As a result, proposing effective methods to control the electronic transport properties of kagome materials has become an urgent research priority. In this context, an exciting advancement came in 2022 when Sun \textit{et al.} successfully isolated a monolayer of $\rm Nb_3Cl_8$ via mechanical exfoliation \cite{acs.nanolett.2c00778}. The three Nb atoms in this material spontaneously form a breathing kagome lattice, where electron cloud overlap is notably stronger within the unit cell than between unit cells. This structural feature breaks the sixfold rotational symmetry, leading to the opening of the Dirac cone and the introduction of multiple band gaps. Building on this, Zhang \textit{et al.} synthesized $\rm Nb_3TeCl_7$ by substituting $\rm Cl$ with $\rm Te$ in $\rm Nb_3Cl_8$ \cite{adma.202301790}, demonstrating that modulating the element at the Te site can tune the breathing effect strength and continuously adjust the band structure. These findings highlight a novel strategy for regulating the metal-insulator transition and magnetic properties, underscoring the importance of investigating the combined roles of breathing and correlation effects in kagome lattices.

To address these research goals, we utilized the DQMC method to study the transport and magnetic properties of the Hubbard model on a breathing kagome lattice. This approach enables precise exploration of the relationship between quantum critical points and the breathing effect, while clarifying underlying physical mechanisms. The paper is structured as follows: Section 2 details the Hubbard Hamiltonian and key observables, with a focus on the DQMC method; Section 3 presents results showing that the breathing effect, in conjunction with interaction strength, drives the kagome system from a paramagnetic metal to a Mott insulator, as evidenced by calculations of kinetic energy, dc conductivity, and density of states, and constructs a phase diagram using correlation and breathing strengths; Section 4 summarizes the main conclusions that the breathing effect can be used to control the band gaps, conductivity, and magnetic properties of kagome materials with electronic interactions.

\begin{figure}[t]
\centering
\includegraphics[width=\linewidth]{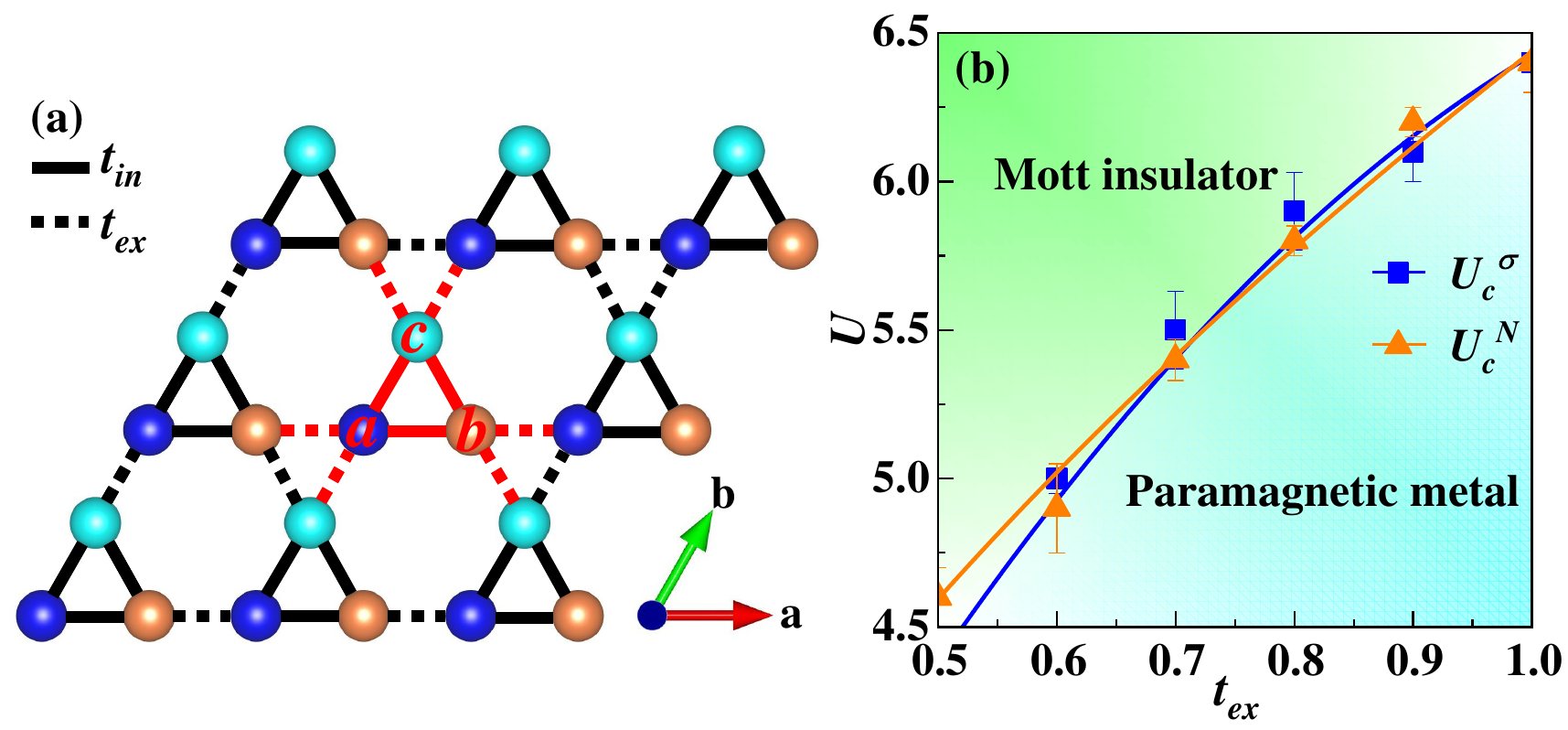}
\caption{(a) The lattice diagram of the breathing kagome system, with the unit cell and hopping processes highlighted in red. (b) The blue and orange curves correspond to the critical interaction strengths \( U_c^\sigma \) (\( U_c^{N} \)) for the metal-insulator transition determined by the dc conductivity \( \sigma_{dc}(T) \) and by the density of states at the Fermi level \( N(0) \), respectively. The green region above the \( U_c^\sigma \) curve represents the Mott insulating phase, while the blue region below the \( U_c^\sigma \) curve represents the paramagnetic metallic phase. Error bars are omitted when they are smaller than the symbol sizes.}
\label{Fig1.pdf}
\end{figure}

\vspace{-0.2cm}  
\section{II. MODEL AND METHOD}
\vspace{-0.3cm}

The Hamiltonian for a Hubbard model on an breathing kagome lattice is defined as
\begin{equation}
\begin{aligned}
H = -\sum_{\mathbf{r},\sigma} \Big\{ & t_{in} \left[a_{\mathbf{r},\sigma}^\dagger b_{\mathbf{r},\sigma} + b_{\mathbf{r},\sigma}^\dagger c_{\mathbf{r},\sigma} + c_{\mathbf{r},\sigma}^\dagger a_{\mathbf{r},\sigma} \right] \\
+ &t_{ex} \left[ a_{\mathbf{r},\sigma}^\dagger b_{\mathbf{r}-\mathbf{x},\sigma} + b_{\mathbf{r},\sigma}^\dagger c_{\mathbf{r}+\mathbf{x}-\mathbf{y},\sigma} + c_{\mathbf{r},\sigma}^\dagger a_{\mathbf{r}+\mathbf{y},\sigma} \right] \Big\} \\
- &\mu \sum_{\mathbf{r},\sigma} n_{\mathbf{r},\sigma} + U \sum_{\mathbf{r}} n_{\mathbf{r},\uparrow} n_{\mathbf{r},\downarrow} + h.c.
\end{aligned}
\label{Eq.1}
\end{equation}
Here, $a_{\mathbf{r},\sigma}$($a_{\mathbf{r},\sigma}^{\dagger}$), $b_{\mathbf{r},\sigma}$($b_{\mathbf{r},\sigma}^{\dagger}$) and $c_{\mathbf{r},\sigma}$($c_{\mathbf{r},\sigma}^{\dagger}$) are the standard fermion annihilation (creation) operators at position $\mathbf{r}$ with spin $\sigma$, and $n_{\mathbf{r},\sigma}$ is the number operator. $t_{in}$, $t_{ex}$, $\mu$, and $U$ represent the hopping amplitude within the unit cell, the hopping amplitude between the unit cells, the chemical potential, and the on-site repulsive interaction, respectively. The sketch for the system that we simulated is shown in Fig. \hyperref[Fig1.pdf]{1(a)}. To introduce the breathing effect in the lattice, we set $t_{in} = 1$ as the reference energy scale and vary $t_{ex}$ from 0.5 to 1 in increments of 0.1 \cite{acs.nanolett.2c00778}. At $t_{ex} = 1$, the system equivalents to a normal kagome lattice. As the value of \( t_{ex} \) decreases, the breathing strength is enhanced.
\par

The DQMC method \cite{PhysRevD.24.2278,PhysRevB.40.506} was utilized to numerically investigate the metal-insulator transition within the Hubbard model as defined by Eq. \hyperref[Eq.1]{1}. As a nonperturbative approach, DQMC provides an exact numerical technique for examining the Hubbard model at finite temperatures. Initially, the partition function \( Z = Tre^{-\beta H} \) is considered as a path integral, which is discretized into \( \Delta \tau \) slices within the imaginary time interval \( (0, \beta) \). The inverse temperature is given by \( \beta \equiv 1/(k_B T) \), with \( k_B \) denoting the Boltzmann constant. The kinetic energy term is quadratic, and the on-site interaction term can be decoupled into a quadratic form by employing a discrete Hubbard-Stratonovich (HS) field transformation. Subsequently, the partition function \( Z \) can be analytically transformed into the product of two fermion determinants by integrating the quadratic Hamiltonian term. These determinants correspond to the spin-up and spin-down states, respectively. To facilitate the sampling process, the Metropolis algorithm is employed with the time-slice interval \( \Delta \tau \) set to 0.1, ensuring sufficiently small errors in the Trotter approximation. To assess the reliability of the computational results, we calculated the average fermion sign \( \langle \text{sign} \rangle \) as a measure of the severity of the sign problem,
\begin{equation}
\langle S \rangle = \frac{\sum_\mathcal{X} det M_{\uparrow}(\mathcal{X}) det M_{\downarrow}(\mathcal{X})}{\sum_{\mathcal{X}} |det M_{\uparrow}(\mathcal{X}) det M_{\downarrow}(\mathcal{X})|},
    \label{Eq.2}
\end{equation}
Here, \( \mathcal{X} \) represents the HS configurations, which consist of spatial sites and imaginary time slices. The matrix \( M_{\sigma}(\mathcal{X}) \) corresponds to the spin species. When \( \langle \text{sign} \rangle = 1 \), the system is free of the sign problem.
\par

To investigate the metal-insulator transition in the breathing kagome lattice, we computed the temperature-dependent dc conductivity. This quantity can be derived from the current-current correlation function, which is momentum \( \mathbf{q} \)-dependent and imaginary time \( \tau \)-dependent, denoted as \( \Lambda_{xx}(\mathbf{q}, \tau) \) \cite{PhysRevB.54.R3756}:
\begin{equation}
\sigma_{dc}(T) = \frac{\beta^2}{\pi} \Lambda_{xx}(\mathbf{q}=0, \tau=\frac{\beta}{2}).
\label{Eq.3}
\end{equation}
Here, $\beta = 1/T$, $\Lambda_{xx}(\mathbf{q}, \tau) = \langle j_x(\mathbf{q}, \tau) j_x(-\mathbf{q}, 0) \rangle$, where $j_x(\mathbf{q}, \tau)$ is the Fourier transform of the unequal-time current-current correlation functions
\begin{equation}
j_x(\mathbf{q}, \tau) = e^{H\tau/\hbar} j_x(\mathbf{r}) e^{-H\tau/\hbar}.
\label{Eq.4}
\end{equation}
The validity of Eq. \hyperref[Eq.3]{3} has been widely established in numerous studies of the metal-insulator transition using the Hubbard model \cite{PhysRevB.105.045132,PhysRevB.109.045107,PhysRevB.101.245161,PhysRevLett.75.312,PhysRevB.90.094506,PhysRevB.107.245126}.
\par

\begin{figure}[t]
	\centering
	\includegraphics[width=\linewidth]{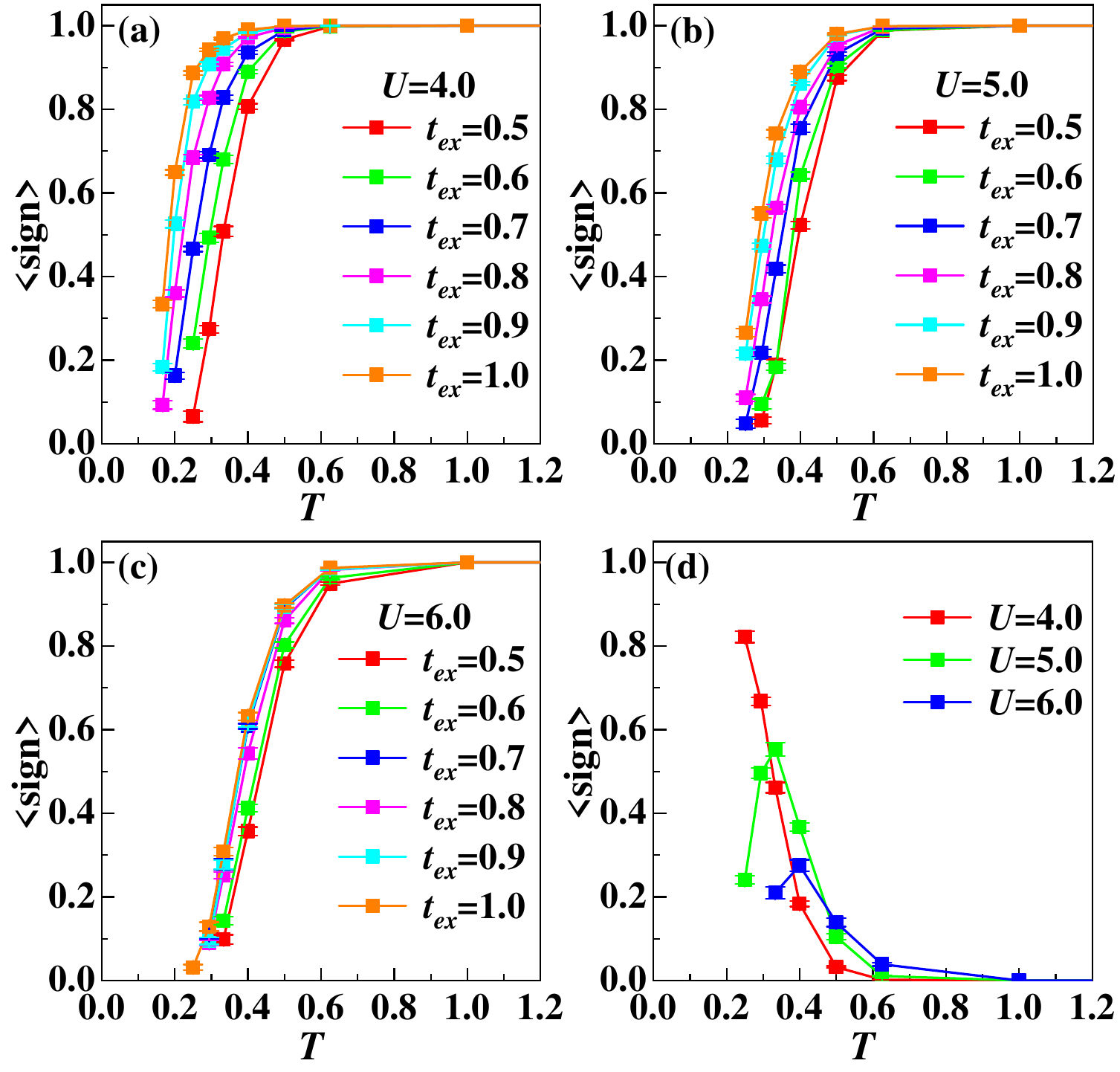}
	\caption{(a)-(c) The average sign \( \langle \text{sign} \rangle \) as a function of temperature \( T \) for \( U = 4.0 \), \( U = 5.0 \), and \( U = 6.0 \). (d) The differences in the average sign \( \langle \text{sign} \rangle \) between \( t_{ex} = 1.0 \) and \( t_{ex} = 0.5 \) for various \( U \) values, denoted as \( \Delta \langle \text{sign} \rangle \). Error bars are omitted when they are smaller than the symbol sizes.}
	\label{Fig2.pdf}
\end{figure}

We also calculated the kinetic energy per site, which can be readily obtained by averaging the hopping terms:
\begin{equation}
\begin{aligned}
\langle K \rangle =& -\frac{1}{N} \sum_{\mathbf{r},\sigma} \bigg\{  t_{in} \left[a_{\mathbf{r},\sigma}^\dagger b_{\mathbf{r},\sigma} + b_{\mathbf{r},\sigma}^\dagger c_{\mathbf{r},\sigma} + c_{\mathbf{r},\sigma}^\dagger a_{\mathbf{r},\sigma} \right] \\
 + &t_{ex} \left[ a_{\mathbf{r},\sigma}^\dagger b_{\mathbf{r}-\mathbf{x},\sigma} + b_{\mathbf{r},\sigma}^\dagger c_{\mathbf{r}+\mathbf{x}-\mathbf{y},\sigma} + c_{\mathbf{r},\sigma}^\dagger a_{\mathbf{r}+\mathbf{y},\sigma} \right] + h.c. \bigg\}.
\end{aligned}
\label{Eq.5}
\end{equation}
To avoid complex numerical analytical continuation methods, the investigation of the density of states is focused on the Fermi level,
\begin{equation}
N(0) \approx \beta \times G(\mathbf{r} = 0, \tau = \beta/2).
\label{Eq.6}
\end{equation}
Here, \( G(\mathbf{r}, \tau) \) is the imaginary-time dependent Green's function. To investigate the magnetic properties of the breathing kagome lattice, we calculated the real-space spin-spin correlation functions,
\begin{equation}
c^{\alpha\beta}(\mathbf{r}) = \frac{1}{3} \langle \mathbf{S}_{\mathbf{r}_0}^\alpha \cdot \mathbf{S}_{\mathbf{r}_0+r}^{\beta} \rangle,
\label{Eq.7}
\end{equation}
where \( \mathbf{r}_0 \) denotes the position of any lattice site, while \( \mathbf{r} \) represents the distance between two sites. The indices \( \alpha \) and \( \beta \) label the sites \( a \), \( b \) or \( c \). After performing a Fourier transform on \( c^{\alpha\beta}(\mathbf{r}) \), we obtain the magnetic structure factor
\begin{equation}
S(\mathbf{q}) = \frac{1}{N_s} \sum_{\alpha,\beta} \sum_{\mathbf{r}} c^{\alpha\beta}(\mathbf{r}) e^{i\mathbf{q*r}},
\label{Eq.8}
\end{equation}
in which \( N_s = 3L^2 \) represents the total number of sites and \( L \) denotes the linear size of the lattice. Furthermore, the uniform magnetic susceptibility is directly related to the magnetic structure factor,
\begin{equation}
\chi \equiv \beta S(0,0).
\label{Eq.9}
\end{equation}

\begin{figure}[t]
\centering
\includegraphics[width=\linewidth]{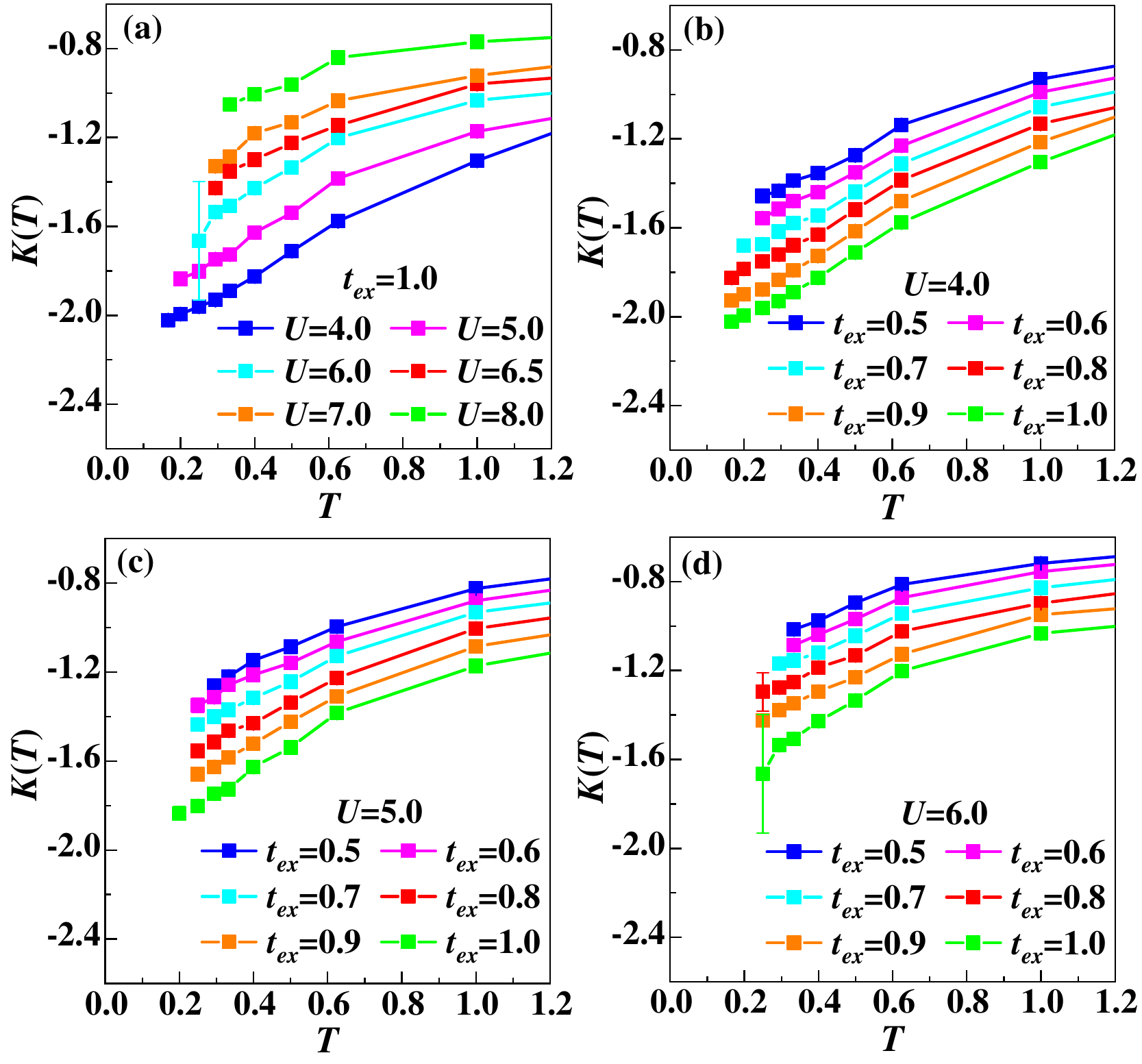}
\caption{(a) The kinetic energy per site \( \langle K \rangle \) for the normal kagome lattice (\( t_{ex} = 1.0 \)) as a function of temperature \( T \) for various \( U \) values. (b)-(d) The kinetic energy per site \( \langle K \rangle \) for the breathing kagome lattice with \( U = 4.0 \), \( U = 5.0 \), and \( U = 6.0 \). Error bars are omitted when they are smaller than the symbol sizes.}
\label{Fig3.pdf}
\end{figure}

\vspace{-0.2cm}
\section{III. RESULT AND DISCUSSION}
\vspace{-0.3cm}

DQMC is an unbiased method, but it is plagued by the notorious minus-sign problem, which results in noisy averages \cite{PhysRevB.41.9301, PhysRevLett.94.170201}. This issue is absent in half-filled systems that possess particle-hole symmetry (PHS), such as bipartite lattices like square and honeycomb lattices. However, the kagome lattice is non-bipartite and lacks PHS for any filling, leading to a severe sign problem that depends on the system size, temperature scale, and interaction strength. 
In present work, we investigated this system at half-filling with the average electron density \( n = 1 \), which represents a regime of significant interest.
To determine the proper monte Carlo parameters rely on the sign problem, we firstly present the average sign \( \langle \text{sign} \rangle \) as a function of temperature \( T \) on the breathing kagome lattice for various \( U \) values in Figs. \hyperref[Fig2.pdf]{2(a)-(c)}. For the same \( T \), \( \langle \text{sign} \rangle \) is significantly suppressed as \( t_{ex} \) decreases, indicating that a larger breathing effect greatly exacerbates the sign problem. Figure \hyperref[Fig2.pdf]{2(d)} shows the differences in \( \langle \text{sign} \rangle \) between \( t_{ex} = 1.0 \) and \( t_{ex} = 0.5 \) for three \( U \) values, defined as \( \Delta \langle \text{sign} \rangle = \langle \text{sign} \rangle_{t_{ex}=1.0} - \langle \text{sign} \rangle_{t_{ex}=0.5} \). When \( U = 4 \), \(\Delta \langle \text{sign} \rangle\) continuously increases as \( T \) decreases. However, when \( U \) is increased to values above 5, \(\Delta \langle \text{sign} \rangle\) first increases to a peak and then begins to decrease. At high temperatures, the average sign \( \langle \text{sign} \rangle \) remains close to 1, ensuring reliable numerical data. As temperature decreases, \( \langle \text{sign} \rangle \) diminishes, reflecting the worsening sign problem. When \( \langle \text{sign} \rangle > 0.5 \), the sign problem remains mild and permits accurate results. For severe cases where \( \langle \text{sign} \rangle < 0.5 \), extended Monte Carlo runs compensate for increased fluctuations within the DQMC framework. We therefore implement substantially longer simulations when \( \langle \text{sign} \rangle \) falls below 0.5 \cite{PhysRevD.24.2278,PhysRevB.94.075106}.

To investigate the effects of the breathing effect and correlated interactions on the transport properties of the kagome lattice, we examine the kinetic energy. Figure \hyperref[Fig3.pdf]{3(a)} displays the kinetic energy per site \( \langle K \rangle \) for the normal kagome lattice for various \( U \) values. The \( \langle K \rangle \) increases with increasing \( T \) and \( U \), suggesting that fermions tend to localize and the system gradually approaches an insulating state. However, up to \( U = 8 \), the system still does not exhibit \( \frac{\partial K(T)}{\partial T} < 0 \). Thus, the behavior of the kinetic energy only indicates a tendency for the kagome lattice to transition towards an insulator, consistent with previous investigations \cite{PhysRevB.107.035134}. Figures \hyperref[Fig3.pdf]{3(b)-(d)} show the kinetic energy including the breathing effect for various \( U \) values. It is evident that \( \langle K \rangle \) increases monotonically with decreasing \( t_{ex} \), but \( \frac{\partial K(T)}{\partial T} < 0 \) still does not appear. Therefore, \( \langle K \rangle \) does not reflect the metal-insulator transition of this system.
\par

\begin{figure}[t]
	\centering
	\includegraphics[width=\linewidth]{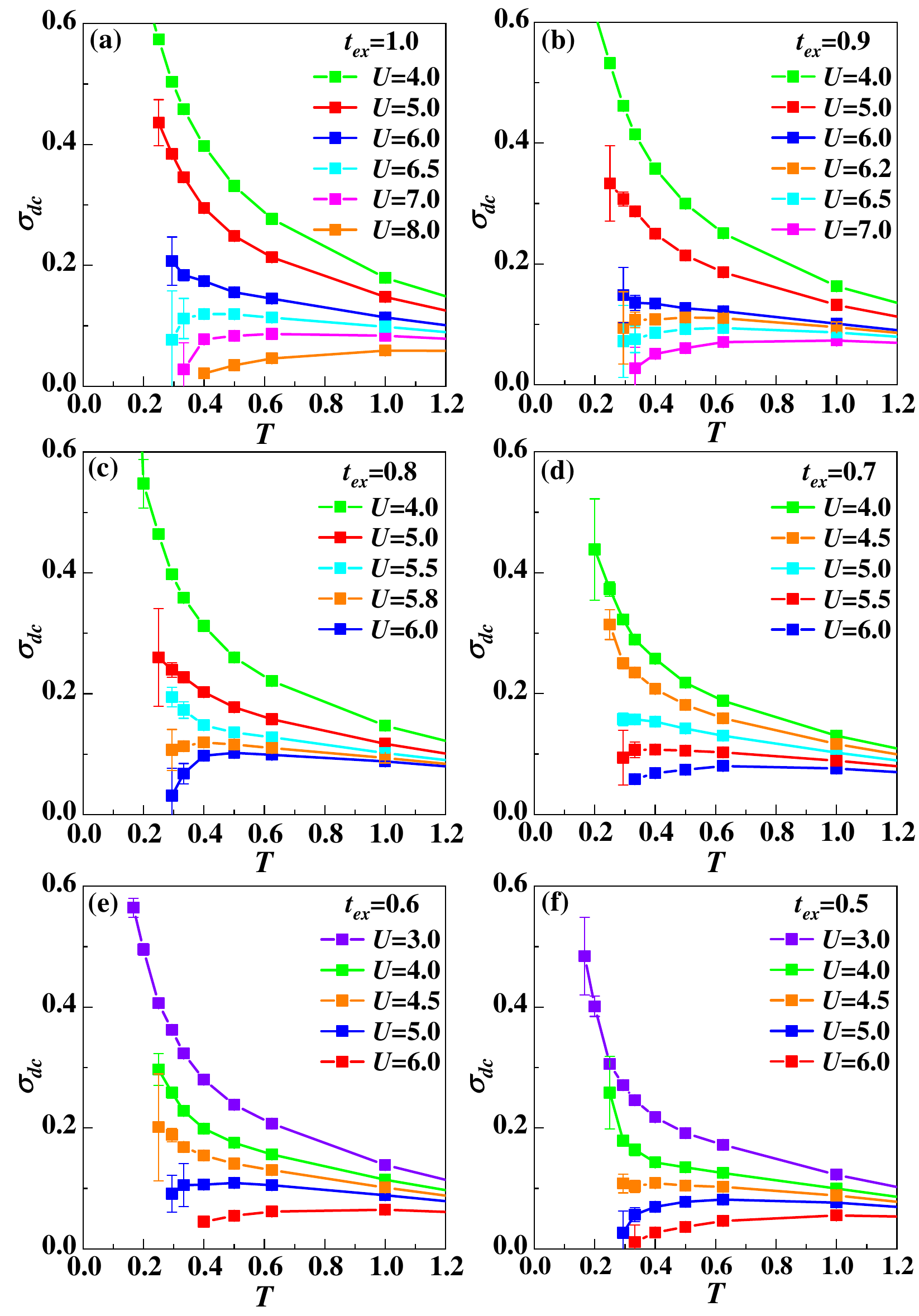}
	\caption{(a)-(f) The dc conductivity \( \sigma_{dc} \) as a function of temperature \( T \) for \( t_{ex} = 1.0 \), \( t_{ex} = 0.9 \), \( t_{ex} = 0.8 \), \( t_{ex} = 0.7 \), \( t_{ex} = 0.6 \), and \( t_{ex} = 0.5 \). Error bars are omitted when they are smaller than the symbol sizes.}
	\label{Fig4.pdf}
\end{figure}

\begin{figure}[t]
	\centering
	\includegraphics[width=\linewidth]{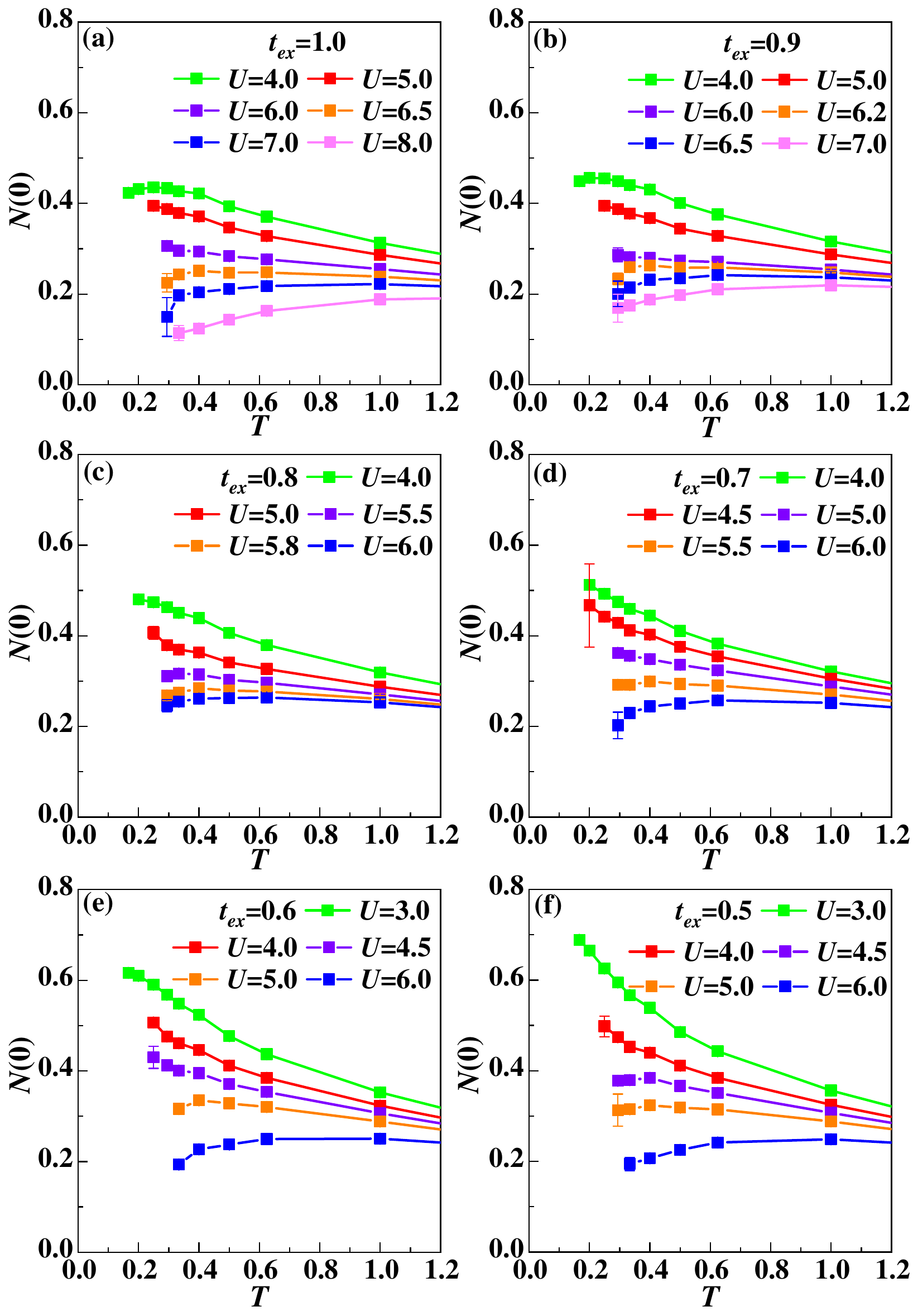}
	\caption{(a)-(f) The density of states at the Fermi level, \( N(0) \), as a function of temperature \( T \) for \( t_{ex} = 1.0 \), \( t_{ex} = 0.9 \), \( t_{ex} = 0.8 \), \( t_{ex} = 0.7 \), \( t_{ex} = 0.6 \), and \( t_{ex} = 0.5 \). Error bars are omitted when they are smaller than the symbol sizes.}
	\label{Fig5.pdf}
\end{figure}

Given the above situation, other quantities should be investigated to identify the phase transition. Therefore, we proceed to examine the dc conductivity \( \sigma_{dc} \) as a function of temperature \( T \), computed across several representative sets of \( U \) and \( t_{ex} \) as shown in Fig. \hyperref[Fig4.pdf]{4}. As illustrated in Fig. \hyperref[Fig4.pdf]{4(a)}, when \( U < 6.5 \), \( \frac{\partial \sigma_{dc}(T)}{\partial T} < 0 \) as \( T \) increases, and \( \sigma_{dc}(T) \) diverges as \( T \) approaches 0, indicating that the system exhibits metallic behavior. However, when \( U \geq 6.5 \), \( \frac{\partial \sigma_{dc}(T)}{\partial T} > 0 \) at low \( T \), which is a signature of fermion localization. This indicates that the system transitions from a metal to an insulator. In this case, we determine the critical interaction strength for the metal-insulator transition defined by the conductivity as \( U_c^\sigma = 6.40 \pm 0.10 \) for \( t_{ex} = 1.0 \). These results are consistent with the preceding study about the normal kagome lattice \cite{PhysRevB.107.035134}. Figures \hyperref[Fig4.pdf]{4(b)-(f)} show the \( \sigma_{dc}(T) \) for various values of \( t_{ex} \). Using the same criterion to determine \( U_c^\sigma \), we find that \( U_c^\sigma \) decreases monotonically with decreasing \( t_{ex} \). Based on these results, we determine the boundary between the metallic and insulating phases, denoted by \( U_c^\sigma \), as shown in Fig. \hyperref[Fig1.pdf]{1(b)}. These findings indicate that the breathing effect reduces the critical interaction strength required for the metal-insulator transition in the correlated kagome system.
\par

\begin{figure}[t]
	\centering
	\includegraphics[width=\linewidth]{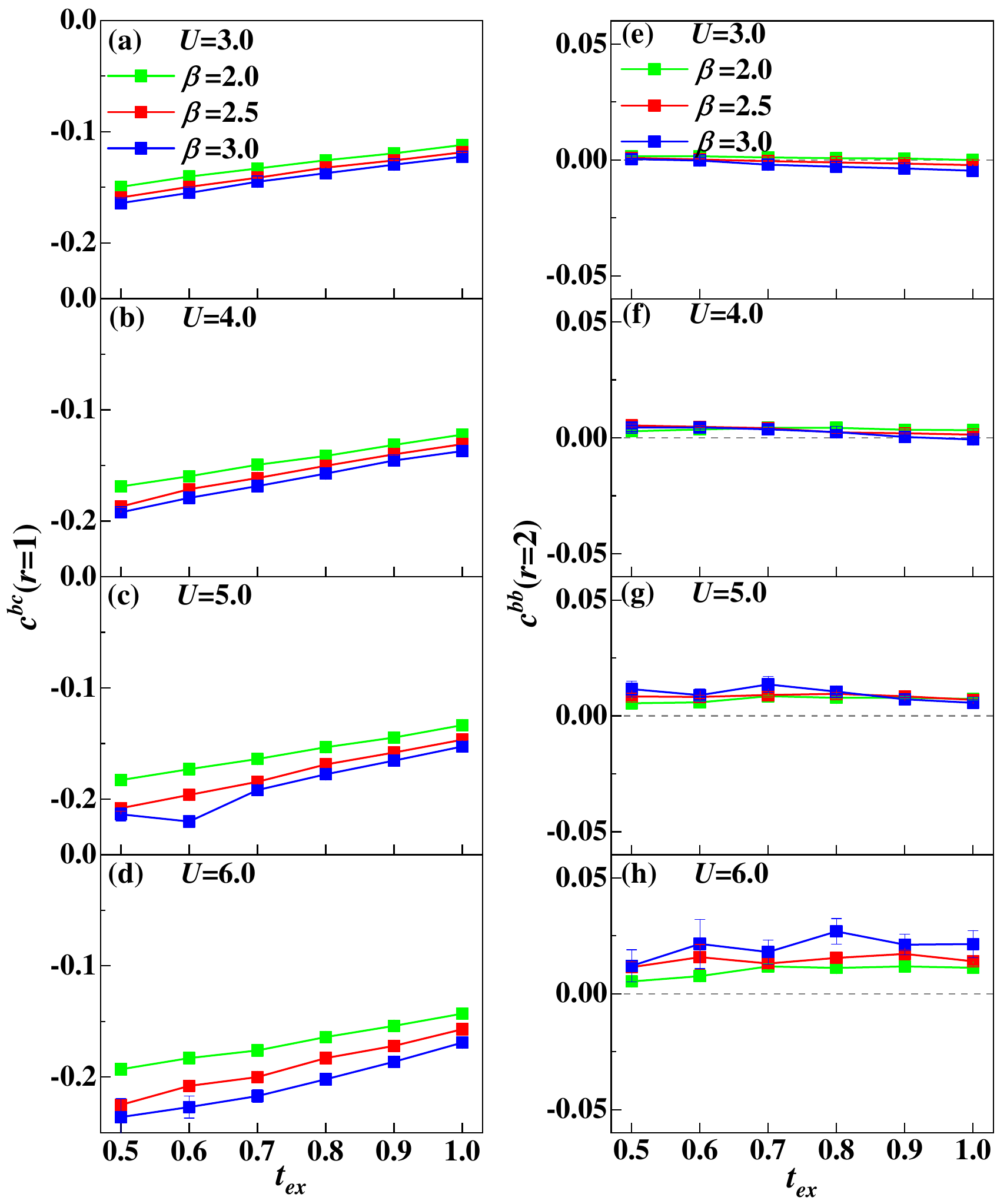}
	\caption{(a)-(d) The spin-spin correlations on the nearest \( b \) and \( c \) sites, \( c^{bc}(r=1) \), for \( U = 3.0 \), \( U = 4.0 \), \( U = 5.0 \), and \( U = 6.0 \). (e)-(h) The spin-spin correlations on the next-nearest \( b \) sites, \( c^{bb}(r=2) \), for \( U = 3.0 \), \( U = 4.0 \), \( U = 5.0 \), and \( U = 6.0 \). Error bars are omitted when they are smaller than the symbol sizes.}
	\label{Fig6.pdf}
\end{figure}

We have analyzed the role of different terms in Eq. (\ref{Eq.1}) in driving the metal-insulator transition by examining $\sigma_{dc}(T)$. This analysis is further corroborated by the behavior of the density of states at the Fermi level, $N(0)$, as $t_{ex}$ is varied. Figure \hyperref[Fig5.pdf]{5(a)} presents $N(0)$ for $t_{ex}=1.0$. For $U < 6.5$, the $\frac{\partial N(0,T)}{\partial T} < 0$ and $N(0,T)$ diverges in the limit as $T$ tends to 0. This behavior is characteristic of a metallic phase. In contrast, when $U \geq 6.5$, the derivative $\frac{\partial N(0,T)}{\partial T}$ becomes positive at low temperatures, indicating a transition to an insulating phase. Thus, the critical interaction strength for the metal-insulator transition, as determined by $N(0)$, is $U_c^N = 6.40 \pm 0.10$ for $t_{ex} = 1.0$. Figures \hyperref[Fig5.pdf]{5(b)-(f)} show the $N(0)$ for various values of \( t_{ex} \). Using the same methodology, we have analyzed the $U_c^N$ for various values of $t_{ex}$, and plotted a boundary separating the metallic and insulating phases, as shown in Fig. \hyperref[Fig1.pdf]{1(b)}. This boundary is in excellent agreement with the metal-insulator transition boundary obtained from the $\sigma_{dc}(T)$ analysis.
\par

\begin{figure}[t]
	\centering
	\includegraphics[width=\linewidth]{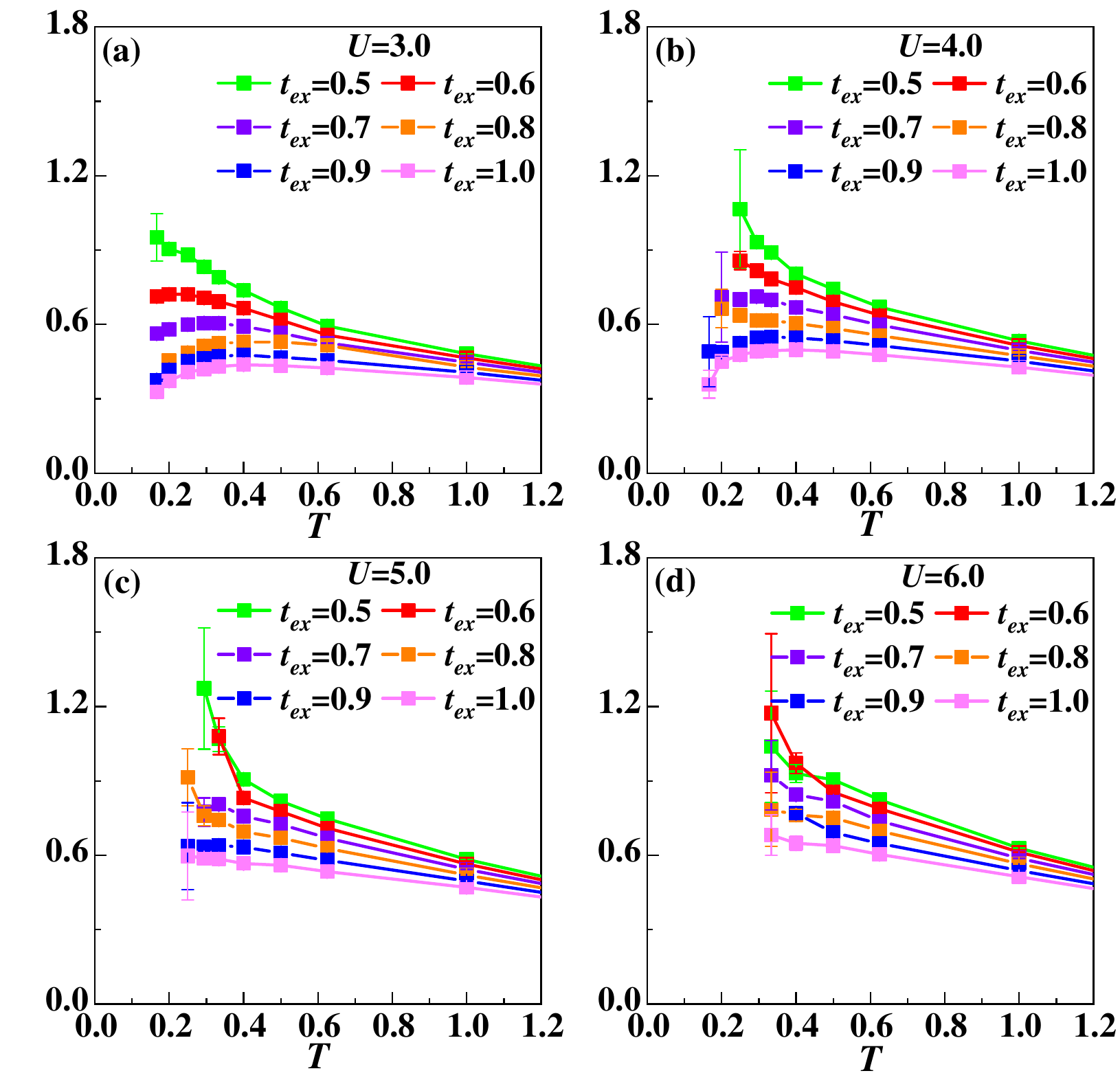}
	\caption{(a)-(d) The temperature dependence of the uniform susceptibility \( \chi \) for \( U = 3.0 \), \( U = 4.0 \), \( U = 5.0 \), and \( U = 6.0 \), as \( t_{ex} \) decreases from 1.0 to 0.5. Error bars are omitted when they are smaller than the symbol sizes.}
	\label{Fig7.pdf}
\end{figure}

To investigate the magnetic properties of the breathing kagome lattice, we examine the behavior of the spin-spin correlation function $c^{bc}(r=1)$ between nearest $b$ and $c$ sites as the $t_{ex}$ decreased for various $U$ (see Figs. \hyperref[Fig6.pdf]{6(a)-(d)}. For all values of $U$ and temperature $T$, the correlation function $c^{bc}(r=1)$ monotonically decreases from negative values as $t_{ex}$ decreased. When $t_{ex}$ and $U$ are held constant, the magnitude of $c^{bc}(r=1)$ increases as the temperature decreases. This indicates that the emergence of magnetic order at low temperatures. Similarly, for a fixed $t_{ex}$ and $T$, the magnitude of $c^{bc}(r=1)$ increases with increasing $U$. This trend reflects that stronger repulsive interactions favor electron localization, thereby enhancing magnetic ordering. The spin-spin correlation function $c^{bb}(r=2)$ for next-nearest neighbors between $b$ sites is depicted in Figs. \hyperref[Fig6.pdf]{6(e)-(h)}. We observe that $c^{bb}(r=2)$ is at least one order of magnitude smaller than $c^{bc}(r=1)$. For $U=3.0$ and $U=4.0$, as shown in Figs. \hyperref[Fig6.pdf]{6(e)-(f)}, the decrease of $t_{ex}$ causes the longer-range correlations to shift slightly towards positive values. When $U$ exceeds 5.0, $c^{bb}(r=2)$ gradually evolves into a weak FM correlation. These observations lead us to conclude that the kagome lattice exhibits strong short-range AFM correlations, which are enhanced as $t_{ex}$ decreases and $U$ increases. Meanwhile, the system haves an extremely weak long-range FM correlation at low $t_{ex}$ and high $U$. Overall, the AFM state consistently dominates the system, regardless of the values of $t_{ex}$ and $U$. This conclusion is consistent with most of the experimentally discovered kagome materials, including $\rm ZnCu_3(OH)_6Cl_2$, $\rm FeGe$, and $\rm YMn_6Sn_6$ \cite{PhysRevLett.98.107204,Nature.022.05034,sciadv.abe2680,PhysRevB.103.014416}.
\par

The above conclusions are further substantiated by the temperature dependence of the uniform susceptibility $\chi$ for various $U$. As shown in Fig. \ref{Fig7.pdf}, the $\chi$ increases as the $t_{ex}$ decreases and $U$ increases, but they do not reach a divergent level. This indicates that the system consistently displays AFM correlations for any values of $t_{ex}$ and $U$. Combining the analyses of the metal-insulator transition and magnetic properties, we present a phase diagram in Fig. \hyperref[Fig1.pdf]{1(b)}. For $U$ values above the boundary (\( U_c^\sigma \)), the conductivity of the system is suppressed, and AFM correlations are strong. We therefore designate this region as the Mott insulating phase. Conversely, when $U$ drops below this boundary, the system exhibits metallic behavior while maintaining the AFM state. This region is thus identified as the paramagnetic metallic phase.
\par

\vspace{-0.2cm}
\section{IV. CONCLUSION}
\vspace{-0.3cm}

Using DQMC simulations, we investigated the metal-insulator transition and magnetic properties of the Hubbard model on a kagome lattice with an introduced breathing effect. The breathing effect is realized by modulating the hopping amplitude between lattice sites. The critical interaction strength for the metal-insulator transition is determined jointly by the calculated electronic kinetic energy, dc conductivity, and the density of electronic states at the Fermi level. Moreover, our analysis of the magnetic properties indicates that the kagome lattice consistently exhibits an AFM state, which is enhanced by the breathing effect and the correlation interaction. By integrating the magnetic and transport responses, we constructed a phase diagram parameterized by $U$ and $t_{ex}$. The region above the boundary \( U_c^\sigma \) is identified as a Mott insulating phase, while the region below the boundary is characterized as a paramagnetic metallic phase. Our study provides valuable insights into the mechanisms underlying the metal-insulator transition and magnetic properties in kagome systems. Additionally, the breathing effect could offer a novel strategy for tuning conductivity and magnetism in strongly correlated materials.
\par

\vspace{-0.2cm}
\section{ACKNOWLEDGMENTS}
\vspace{-0.3cm}
This work was supported by NSFC (12474218) and Beijing Natural Science Foundation (No. 1242022 and 1252022). The numerical simulations in this work were performed at the HSCC of Beijing Normal University.

\bibliography{ref}

\begin{thebibliography}{50}%
\makeatletter
\providecommand \@ifxundefined [1]{%
 \@ifx{#1\undefined}
}%
\providecommand \@ifnum [1]{%
 \ifnum #1\expandafter \@firstoftwo
 \else \expandafter \@secondoftwo
 \fi
}%
\providecommand \@ifx [1]{%
 \ifx #1\expandafter \@firstoftwo
 \else \expandafter \@secondoftwo
 \fi
}%
\providecommand \natexlab [1]{#1}%
\providecommand \enquote  [1]{``#1''}%
\providecommand \bibnamefont  [1]{#1}%
\providecommand \bibfnamefont [1]{#1}%
\providecommand \citenamefont [1]{#1}%
\providecommand \href@noop [0]{\@secondoftwo}%
\providecommand \href [0]{\begingroup \@sanitize@url \@href}%
\providecommand \@href[1]{\@@startlink{#1}\@@href}%
\providecommand \@@href[1]{\endgroup#1\@@endlink}%
\providecommand \@sanitize@url [0]{\catcode `\\12\catcode `\$12\catcode
  `\&12\catcode `\#12\catcode `\^12\catcode `\_12\catcode `\%12\relax}%
\providecommand \@@startlink[1]{}%
\providecommand \@@endlink[0]{}%
\providecommand \url  [0]{\begingroup\@sanitize@url \@url }%
\providecommand \@url [1]{\endgroup\@href {#1}{\urlprefix }}%
\providecommand \urlprefix  [0]{URL }%
\providecommand \Eprint [0]{\href }%
\providecommand \doibase [0]{https://doi.org/}%
\providecommand \selectlanguage [0]{\@gobble}%
\providecommand \bibinfo  [0]{\@secondoftwo}%
\providecommand \bibfield  [0]{\@secondoftwo}%
\providecommand \translation [1]{[#1]}%
\providecommand \BibitemOpen [0]{}%
\providecommand \bibitemStop [0]{}%
\providecommand \bibitemNoStop [0]{.\EOS\space}%
\providecommand \EOS [0]{\spacefactor3000\relax}%
\providecommand \BibitemShut  [1]{\csname bibitem#1\endcsname}%
\let\auto@bib@innerbib\@empty
\bibitem [{\citenamefont {Liu}\ \emph {et~al.}(2021)\citenamefont {Liu},
  \citenamefont {Zhao}, \citenamefont {Yin}, \citenamefont {Gong},
  \citenamefont {Tu}, \citenamefont {Li}, \citenamefont {Song}, \citenamefont
  {Liu}, \citenamefont {Shen}, \citenamefont {Huang} \emph
  {et~al.}}]{PhysRevX.11.041010}%
  \BibitemOpen
  \bibfield  {author} {\bibinfo {author} {\bibfnamefont {Z.}~\bibnamefont
  {Liu}}, \bibinfo {author} {\bibfnamefont {N.}~\bibnamefont {Zhao}}, \bibinfo
  {author} {\bibfnamefont {Q.}~\bibnamefont {Yin}}, \bibinfo {author}
  {\bibfnamefont {C.}~\bibnamefont {Gong}}, \bibinfo {author} {\bibfnamefont
  {Z.}~\bibnamefont {Tu}}, \bibinfo {author} {\bibfnamefont {M.}~\bibnamefont
  {Li}}, \bibinfo {author} {\bibfnamefont {W.}~\bibnamefont {Song}}, \bibinfo
  {author} {\bibfnamefont {Z.}~\bibnamefont {Liu}}, \bibinfo {author}
  {\bibfnamefont {D.}~\bibnamefont {Shen}}, \bibinfo {author} {\bibfnamefont
  {Y.}~\bibnamefont {Huang}}, \emph {et~al.},\ }\href
  {https://doi.org/10.1103/PhysRevX.11.041010} {\bibfield  {journal} {\bibinfo
  {journal} {Phys. Rev. X}\ }\textbf {\bibinfo {volume} {11}},\ \bibinfo
  {pages} {041010} (\bibinfo {year} {2021})}\BibitemShut {NoStop}%
\bibitem [{\citenamefont {Cho}\ \emph {et~al.}(2021)\citenamefont {Cho},
  \citenamefont {Ma}, \citenamefont {Xia}, \citenamefont {Yang}, \citenamefont
  {Liu}, \citenamefont {Huang}, \citenamefont {Jiang}, \citenamefont {Lu},
  \citenamefont {Liu}, \citenamefont {Liu} \emph
  {et~al.}}]{PhysRevLett.127.236401}%
  \BibitemOpen
  \bibfield  {author} {\bibinfo {author} {\bibfnamefont {S.}~\bibnamefont
  {Cho}}, \bibinfo {author} {\bibfnamefont {H.}~\bibnamefont {Ma}}, \bibinfo
  {author} {\bibfnamefont {W.}~\bibnamefont {Xia}}, \bibinfo {author}
  {\bibfnamefont {Y.}~\bibnamefont {Yang}}, \bibinfo {author} {\bibfnamefont
  {Z.}~\bibnamefont {Liu}}, \bibinfo {author} {\bibfnamefont {Z.}~\bibnamefont
  {Huang}}, \bibinfo {author} {\bibfnamefont {Z.}~\bibnamefont {Jiang}},
  \bibinfo {author} {\bibfnamefont {X.}~\bibnamefont {Lu}}, \bibinfo {author}
  {\bibfnamefont {J.}~\bibnamefont {Liu}}, \bibinfo {author} {\bibfnamefont
  {Z.}~\bibnamefont {Liu}}, \emph {et~al.},\ }\href
  {https://doi.org/10.1103/PhysRevLett.127.236401} {\bibfield  {journal}
  {\bibinfo  {journal} {Phys. Rev. Lett.}\ }\textbf {\bibinfo {volume} {127}},\
  \bibinfo {pages} {236401} (\bibinfo {year} {2021})}\BibitemShut {NoStop}%
\bibitem [{\citenamefont {Yi}\ \emph {et~al.}(2025)\citenamefont {Yi},
  \citenamefont {Liao}, \citenamefont {Wang}, \citenamefont {Ma}, \citenamefont
  {Liu}, \citenamefont {Teng}, \citenamefont {Gao}, \citenamefont {Dai},
  \citenamefont {Dai}, \citenamefont {Zhao} \emph
  {et~al.}}]{PhysRevLett.134.086902}%
  \BibitemOpen
  \bibfield  {author} {\bibinfo {author} {\bibfnamefont {S.}~\bibnamefont
  {Yi}}, \bibinfo {author} {\bibfnamefont {Z.}~\bibnamefont {Liao}}, \bibinfo
  {author} {\bibfnamefont {Q.}~\bibnamefont {Wang}}, \bibinfo {author}
  {\bibfnamefont {H.}~\bibnamefont {Ma}}, \bibinfo {author} {\bibfnamefont
  {J.}~\bibnamefont {Liu}}, \bibinfo {author} {\bibfnamefont {X.}~\bibnamefont
  {Teng}}, \bibinfo {author} {\bibfnamefont {B.}~\bibnamefont {Gao}}, \bibinfo
  {author} {\bibfnamefont {P.}~\bibnamefont {Dai}}, \bibinfo {author}
  {\bibfnamefont {Y.}~\bibnamefont {Dai}}, \bibinfo {author} {\bibfnamefont
  {J.}~\bibnamefont {Zhao}}, \emph {et~al.},\ }\href
  {https://doi.org/10.1103/PhysRevLett.134.086902} {\bibfield  {journal}
  {\bibinfo  {journal} {Phys. Rev. Lett.}\ }\textbf {\bibinfo {volume} {134}},\
  \bibinfo {pages} {086902} (\bibinfo {year} {2025})}\BibitemShut {NoStop}%
\bibitem [{\citenamefont {Chen}\ \emph
  {et~al.}(2021{\natexlab{a}})\citenamefont {Chen}, \citenamefont {Yang},
  \citenamefont {Hu}, \citenamefont {Zhao}, \citenamefont {Yuan}, \citenamefont
  {Xing}, \citenamefont {Qian}, \citenamefont {Huang}, \citenamefont {Li},
  \citenamefont {Ye} \emph {et~al.}}]{Nature.s41586-021-03983-5}%
  \BibitemOpen
  \bibfield  {author} {\bibinfo {author} {\bibfnamefont {H.}~\bibnamefont
  {Chen}}, \bibinfo {author} {\bibfnamefont {H.}~\bibnamefont {Yang}}, \bibinfo
  {author} {\bibfnamefont {B.}~\bibnamefont {Hu}}, \bibinfo {author}
  {\bibfnamefont {Z.}~\bibnamefont {Zhao}}, \bibinfo {author} {\bibfnamefont
  {J.}~\bibnamefont {Yuan}}, \bibinfo {author} {\bibfnamefont {Y.}~\bibnamefont
  {Xing}}, \bibinfo {author} {\bibfnamefont {G.}~\bibnamefont {Qian}}, \bibinfo
  {author} {\bibfnamefont {Z.}~\bibnamefont {Huang}}, \bibinfo {author}
  {\bibfnamefont {G.}~\bibnamefont {Li}}, \bibinfo {author} {\bibfnamefont
  {Y.}~\bibnamefont {Ye}}, \emph {et~al.},\ }\href
  {https://doi.org/10.1038/s41586-021-03983-5} {\bibfield  {journal} {\bibinfo
  {journal} {Nature}\ }\textbf {\bibinfo {volume} {599}},\ \bibinfo {pages}
  {222} (\bibinfo {year} {2021}{\natexlab{a}})}\BibitemShut {NoStop}%
\bibitem [{\citenamefont {Balents}(2010)}]{nature08917}%
  \BibitemOpen
  \bibfield  {author} {\bibinfo {author} {\bibfnamefont {L.}~\bibnamefont
  {Balents}},\ }\href {https://doi.org/10.1038/nature08917} {\bibfield
  {journal} {\bibinfo  {journal} {Nature}\ }\textbf {\bibinfo {volume} {464}},\
  \bibinfo {pages} {199} (\bibinfo {year} {2010})}\BibitemShut {NoStop}%
\bibitem [{\citenamefont {Han}\ \emph {et~al.}(2012)\citenamefont {Han},
  \citenamefont {Helton}, \citenamefont {Chu}, \citenamefont {Nocera},
  \citenamefont {Rodriguez-Rivera}, \citenamefont {Broholm},\ and\
  \citenamefont {Lee}}]{Nature.11659}%
  \BibitemOpen
  \bibfield  {author} {\bibinfo {author} {\bibfnamefont {T.-H.}\ \bibnamefont
  {Han}}, \bibinfo {author} {\bibfnamefont {J.~S.}\ \bibnamefont {Helton}},
  \bibinfo {author} {\bibfnamefont {S.}~\bibnamefont {Chu}}, \bibinfo {author}
  {\bibfnamefont {D.~G.}\ \bibnamefont {Nocera}}, \bibinfo {author}
  {\bibfnamefont {J.~A.}\ \bibnamefont {Rodriguez-Rivera}}, \bibinfo {author}
  {\bibfnamefont {C.}~\bibnamefont {Broholm}},\ and\ \bibinfo {author}
  {\bibfnamefont {Y.~S.}\ \bibnamefont {Lee}},\ }\href
  {https://doi.org/10.1038/nature11659} {\bibfield  {journal} {\bibinfo
  {journal} {Nature}\ }\textbf {\bibinfo {volume} {492}},\ \bibinfo {pages}
  {406} (\bibinfo {year} {2012})}\BibitemShut {NoStop}%
\bibitem [{\citenamefont {Fu}\ \emph {et~al.}(2015)\citenamefont {Fu},
  \citenamefont {Imai}, \citenamefont {Han},\ and\ \citenamefont
  {Lee}}]{science.aab2120}%
  \BibitemOpen
  \bibfield  {author} {\bibinfo {author} {\bibfnamefont {M.}~\bibnamefont
  {Fu}}, \bibinfo {author} {\bibfnamefont {T.}~\bibnamefont {Imai}}, \bibinfo
  {author} {\bibfnamefont {T.-H.}\ \bibnamefont {Han}},\ and\ \bibinfo {author}
  {\bibfnamefont {Y.~S.}\ \bibnamefont {Lee}},\ }\href
  {https://doi.org/10.1126/science.aab2120} {\bibfield  {journal} {\bibinfo
  {journal} {Science}\ }\textbf {\bibinfo {volume} {350}},\ \bibinfo {pages}
  {655} (\bibinfo {year} {2015})}\BibitemShut {NoStop}%
\bibitem [{\citenamefont {Chen}\ \emph
  {et~al.}(2021{\natexlab{b}})\citenamefont {Chen}, \citenamefont {Wang},
  \citenamefont {Yin}, \citenamefont {Gu}, \citenamefont {Jiang}, \citenamefont
  {Tu}, \citenamefont {Gong}, \citenamefont {Uwatoko}, \citenamefont {Sun},
  \citenamefont {Lei}, \citenamefont {Hu}, \citenamefont {Cheng} \emph
  {et~al.}}]{PhysRevLett.126.247001}%
  \BibitemOpen
  \bibfield  {author} {\bibinfo {author} {\bibfnamefont {K.~Y.}\ \bibnamefont
  {Chen}}, \bibinfo {author} {\bibfnamefont {N.~N.}\ \bibnamefont {Wang}},
  \bibinfo {author} {\bibfnamefont {Q.~W.}\ \bibnamefont {Yin}}, \bibinfo
  {author} {\bibfnamefont {Y.~H.}\ \bibnamefont {Gu}}, \bibinfo {author}
  {\bibfnamefont {K.}~\bibnamefont {Jiang}}, \bibinfo {author} {\bibfnamefont
  {Z.~J.}\ \bibnamefont {Tu}}, \bibinfo {author} {\bibfnamefont {C.~S.}\
  \bibnamefont {Gong}}, \bibinfo {author} {\bibfnamefont {Y.}~\bibnamefont
  {Uwatoko}}, \bibinfo {author} {\bibfnamefont {J.~P.}\ \bibnamefont {Sun}},
  \bibinfo {author} {\bibfnamefont {H.~C.}\ \bibnamefont {Lei}}, \bibinfo
  {author} {\bibfnamefont {J.~P.}\ \bibnamefont {Hu}}, \bibinfo {author}
  {\bibfnamefont {J.-G.}\ \bibnamefont {Cheng}}, \emph {et~al.},\ }\href
  {https://doi.org/10.1103/PhysRevLett.126.247001} {\bibfield  {journal}
  {\bibinfo  {journal} {Phys. Rev. Lett.}\ }\textbf {\bibinfo {volume} {126}},\
  \bibinfo {pages} {247001} (\bibinfo {year} {2021}{\natexlab{b}})}\BibitemShut
  {NoStop}%
\bibitem [{\citenamefont {Hu}\ \emph {et~al.}(2023)\citenamefont {Hu},
  \citenamefont {Le}, \citenamefont {Zhang}, \citenamefont {Zhao},
  \citenamefont {Liu}, \citenamefont {Ma}, \citenamefont {Plumb}, \citenamefont
  {Radovic}, \citenamefont {Chen}, \citenamefont {Schnyder} \emph
  {et~al.}}]{NatPhys.s41567-023-02215-z}%
  \BibitemOpen
  \bibfield  {author} {\bibinfo {author} {\bibfnamefont {Y.}~\bibnamefont
  {Hu}}, \bibinfo {author} {\bibfnamefont {C.}~\bibnamefont {Le}}, \bibinfo
  {author} {\bibfnamefont {Y.}~\bibnamefont {Zhang}}, \bibinfo {author}
  {\bibfnamefont {Z.}~\bibnamefont {Zhao}}, \bibinfo {author} {\bibfnamefont
  {J.}~\bibnamefont {Liu}}, \bibinfo {author} {\bibfnamefont {J.}~\bibnamefont
  {Ma}}, \bibinfo {author} {\bibfnamefont {N.~C.}\ \bibnamefont {Plumb}},
  \bibinfo {author} {\bibfnamefont {M.}~\bibnamefont {Radovic}}, \bibinfo
  {author} {\bibfnamefont {H.}~\bibnamefont {Chen}}, \bibinfo {author}
  {\bibfnamefont {A.~P.}\ \bibnamefont {Schnyder}}, \emph {et~al.},\ }\href
  {https://doi.org/10.1038/s41567-023-02215-z} {\bibfield  {journal} {\bibinfo
  {journal} {Nat. Phys.}\ }\textbf {\bibinfo {volume} {19}},\ \bibinfo {pages}
  {1827} (\bibinfo {year} {2023})}\BibitemShut {NoStop}%
\bibitem [{\citenamefont {Le}\ \emph {et~al.}(2024)\citenamefont {Le},
  \citenamefont {Pan}, \citenamefont {Xu}, \citenamefont {Liu}, \citenamefont
  {Wang}, \citenamefont {Lou}, \citenamefont {Yang}, \citenamefont {Wang},
  \citenamefont {Yao}, \citenamefont {Wu} \emph
  {et~al.}}]{Nature.s41586-024-07431-y}%
  \BibitemOpen
  \bibfield  {author} {\bibinfo {author} {\bibfnamefont {T.}~\bibnamefont
  {Le}}, \bibinfo {author} {\bibfnamefont {Z.}~\bibnamefont {Pan}}, \bibinfo
  {author} {\bibfnamefont {Z.}~\bibnamefont {Xu}}, \bibinfo {author}
  {\bibfnamefont {J.}~\bibnamefont {Liu}}, \bibinfo {author} {\bibfnamefont
  {J.}~\bibnamefont {Wang}}, \bibinfo {author} {\bibfnamefont {Z.}~\bibnamefont
  {Lou}}, \bibinfo {author} {\bibfnamefont {X.}~\bibnamefont {Yang}}, \bibinfo
  {author} {\bibfnamefont {Z.}~\bibnamefont {Wang}}, \bibinfo {author}
  {\bibfnamefont {Y.}~\bibnamefont {Yao}}, \bibinfo {author} {\bibfnamefont
  {C.}~\bibnamefont {Wu}}, \emph {et~al.},\ }\href
  {https://doi.org/10.1038/s41586-024-07431-y} {\bibfield  {journal} {\bibinfo
  {journal} {Nature}\ }\textbf {\bibinfo {volume} {630}},\ \bibinfo {pages}
  {64} (\bibinfo {year} {2024})}\BibitemShut {NoStop}%
\bibitem [{\citenamefont {Acharya}\ \emph {et~al.}(2024)\citenamefont
  {Acharya}, \citenamefont {Neupane}, \citenamefont {Hsu}, \citenamefont
  {Yang}, \citenamefont {Graf}, \citenamefont {Choi}, \citenamefont {Pandey},
  \citenamefont {Nabi}, \citenamefont {Chhetri}, \citenamefont {Basnet} \emph
  {et~al.}}]{AdvMater.adma.202410655}%
  \BibitemOpen
  \bibfield  {author} {\bibinfo {author} {\bibfnamefont {G.}~\bibnamefont
  {Acharya}}, \bibinfo {author} {\bibfnamefont {B.}~\bibnamefont {Neupane}},
  \bibinfo {author} {\bibfnamefont {C.-H.}\ \bibnamefont {Hsu}}, \bibinfo
  {author} {\bibfnamefont {X.~P.}\ \bibnamefont {Yang}}, \bibinfo {author}
  {\bibfnamefont {D.}~\bibnamefont {Graf}}, \bibinfo {author} {\bibfnamefont
  {E.~S.}\ \bibnamefont {Choi}}, \bibinfo {author} {\bibfnamefont
  {K.}~\bibnamefont {Pandey}}, \bibinfo {author} {\bibfnamefont {M.~R.~U.}\
  \bibnamefont {Nabi}}, \bibinfo {author} {\bibfnamefont {S.~K.}\ \bibnamefont
  {Chhetri}}, \bibinfo {author} {\bibfnamefont {R.}~\bibnamefont {Basnet}},
  \emph {et~al.},\ }\href {https://doi.org/10.1002/adma.202410655} {\bibfield
  {journal} {\bibinfo  {journal} {Adv. Mater.}\ }\textbf {\bibinfo {volume}
  {36}},\ \bibinfo {pages} {2407655} (\bibinfo {year} {2024})}\BibitemShut
  {NoStop}%
\bibitem [{\citenamefont {Teng}\ \emph {et~al.}(2022)\citenamefont {Teng},
  \citenamefont {Chen}, \citenamefont {Ye}, \citenamefont {Rosenberg},
  \citenamefont {Liu}, \citenamefont {Yin}, \citenamefont {Jiang},
  \citenamefont {Oh}, \citenamefont {Hasan}, \citenamefont {Neubauer} \emph
  {et~al.}}]{Nature.022.05034}%
  \BibitemOpen
  \bibfield  {author} {\bibinfo {author} {\bibfnamefont {X.}~\bibnamefont
  {Teng}}, \bibinfo {author} {\bibfnamefont {L.}~\bibnamefont {Chen}}, \bibinfo
  {author} {\bibfnamefont {F.}~\bibnamefont {Ye}}, \bibinfo {author}
  {\bibfnamefont {E.}~\bibnamefont {Rosenberg}}, \bibinfo {author}
  {\bibfnamefont {Z.}~\bibnamefont {Liu}}, \bibinfo {author} {\bibfnamefont
  {J.-X.}\ \bibnamefont {Yin}}, \bibinfo {author} {\bibfnamefont {Y.-X.}\
  \bibnamefont {Jiang}}, \bibinfo {author} {\bibfnamefont {J.~S.}\ \bibnamefont
  {Oh}}, \bibinfo {author} {\bibfnamefont {M.~Z.}\ \bibnamefont {Hasan}},
  \bibinfo {author} {\bibfnamefont {K.~J.}\ \bibnamefont {Neubauer}}, \emph
  {et~al.},\ }\href {https://doi.org/10.1038/s41586-022-05034-z} {\bibfield
  {journal} {\bibinfo  {journal} {Nature}\ }\textbf {\bibinfo {volume} {609}},\
  \bibinfo {pages} {490} (\bibinfo {year} {2022})}\BibitemShut {NoStop}%
\bibitem [{\citenamefont {Sun}\ \emph {et~al.}(2022)\citenamefont {Sun},
  \citenamefont {Zhou}, \citenamefont {Wang}, \citenamefont {Kumar},
  \citenamefont {Geng}, \citenamefont {Yue}, \citenamefont {Han}, \citenamefont
  {Haraguchi}, \citenamefont {Shimada}, \citenamefont {Cheng} \emph
  {et~al.}}]{acs.nanolett.2c00778}%
  \BibitemOpen
  \bibfield  {author} {\bibinfo {author} {\bibfnamefont {Z.}~\bibnamefont
  {Sun}}, \bibinfo {author} {\bibfnamefont {H.}~\bibnamefont {Zhou}}, \bibinfo
  {author} {\bibfnamefont {C.}~\bibnamefont {Wang}}, \bibinfo {author}
  {\bibfnamefont {S.}~\bibnamefont {Kumar}}, \bibinfo {author} {\bibfnamefont
  {D.}~\bibnamefont {Geng}}, \bibinfo {author} {\bibfnamefont {S.}~\bibnamefont
  {Yue}}, \bibinfo {author} {\bibfnamefont {X.}~\bibnamefont {Han}}, \bibinfo
  {author} {\bibfnamefont {Y.}~\bibnamefont {Haraguchi}}, \bibinfo {author}
  {\bibfnamefont {K.}~\bibnamefont {Shimada}}, \bibinfo {author} {\bibfnamefont
  {P.}~\bibnamefont {Cheng}}, \emph {et~al.},\ }\href
  {https://doi.org/10.1021/acs.nanolett.2c00778} {\bibfield  {journal}
  {\bibinfo  {journal} {Nano Lett.}\ }\textbf {\bibinfo {volume} {22}},\
  \bibinfo {pages} {4596} (\bibinfo {year} {2022})}\BibitemShut {NoStop}%
\bibitem [{\citenamefont {Zhang}\ \emph {et~al.}(2023)\citenamefont {Zhang},
  \citenamefont {Shi}, \citenamefont {Jiang}, \citenamefont {Yang},
  \citenamefont {Zhang}, \citenamefont {Meng}, \citenamefont {Hu},
  \citenamefont {Liu}, \citenamefont {Cheng}, \citenamefont {Xie} \emph
  {et~al.}}]{adma.202301790}%
  \BibitemOpen
  \bibfield  {author} {\bibinfo {author} {\bibfnamefont {H.}~\bibnamefont
  {Zhang}}, \bibinfo {author} {\bibfnamefont {Z.}~\bibnamefont {Shi}}, \bibinfo
  {author} {\bibfnamefont {Z.}~\bibnamefont {Jiang}}, \bibinfo {author}
  {\bibfnamefont {M.}~\bibnamefont {Yang}}, \bibinfo {author} {\bibfnamefont
  {J.}~\bibnamefont {Zhang}}, \bibinfo {author} {\bibfnamefont
  {Z.}~\bibnamefont {Meng}}, \bibinfo {author} {\bibfnamefont {T.}~\bibnamefont
  {Hu}}, \bibinfo {author} {\bibfnamefont {F.}~\bibnamefont {Liu}}, \bibinfo
  {author} {\bibfnamefont {L.}~\bibnamefont {Cheng}}, \bibinfo {author}
  {\bibfnamefont {Y.}~\bibnamefont {Xie}}, \emph {et~al.},\ }\href
  {https://doi.org/10.1002/adma.202301790} {\bibfield  {journal} {\bibinfo
  {journal} {Adv. Mater.}\ }\textbf {\bibinfo {volume} {35}},\ \bibinfo {pages}
  {2301790} (\bibinfo {year} {2023})}\BibitemShut {NoStop}%
\bibitem [{\citenamefont {Lee}\ \emph {et~al.}(2015)\citenamefont {Lee},
  \citenamefont {Richardella}, \citenamefont {Hickey}, \citenamefont
  {Mkhoyan},\ and\ \citenamefont {Samarth}}]{PhysRevB.92.155312}%
  \BibitemOpen
  \bibfield  {author} {\bibinfo {author} {\bibfnamefont {J.~S.}\ \bibnamefont
  {Lee}}, \bibinfo {author} {\bibfnamefont {A.}~\bibnamefont {Richardella}},
  \bibinfo {author} {\bibfnamefont {D.~R.}\ \bibnamefont {Hickey}}, \bibinfo
  {author} {\bibfnamefont {K.~A.}\ \bibnamefont {Mkhoyan}},\ and\ \bibinfo
  {author} {\bibfnamefont {N.}~\bibnamefont {Samarth}},\ }\href
  {https://link.aps.org/doi/10.1103/PhysRevB.92.155312} {\bibfield  {journal}
  {\bibinfo  {journal} {Phys. Rev. B}\ }\textbf {\bibinfo {volume} {92}},\
  \bibinfo {pages} {155312} (\bibinfo {year} {2015})}\BibitemShut {NoStop}%
\bibitem [{\citenamefont {Musfeldt}\ \emph {et~al.}(2023)\citenamefont
  {Musfeldt}, \citenamefont {Mandrus},\ and\ \citenamefont
  {Liu}}]{npj.023.00389}%
  \BibitemOpen
  \bibfield  {author} {\bibinfo {author} {\bibfnamefont {J.~L.}\ \bibnamefont
  {Musfeldt}}, \bibinfo {author} {\bibfnamefont {D.~G.}\ \bibnamefont
  {Mandrus}},\ and\ \bibinfo {author} {\bibfnamefont {Z.}~\bibnamefont {Liu}},\
  }\href {https://doi.org/10.1038/s41699-023-00389-x} {\bibfield  {journal}
  {\bibinfo  {journal} {npj 2D Mater. Appl.}\ }\textbf {\bibinfo {volume}
  {7}},\ \bibinfo {pages} {28} (\bibinfo {year} {2023})}\BibitemShut {NoStop}%
\bibitem [{\citenamefont {Helton}\ \emph {et~al.}(2007)\citenamefont {Helton},
  \citenamefont {Matan}, \citenamefont {Shores}, \citenamefont {Nytko},
  \citenamefont {Bartlett}, \citenamefont {Yoshida}, \citenamefont {Takano},
  \citenamefont {Suslov}, \citenamefont {Qiu}, \citenamefont {Chung} \emph
  {et~al.}}]{PhysRevLett.98.107204}%
  \BibitemOpen
  \bibfield  {author} {\bibinfo {author} {\bibfnamefont {J.~S.}\ \bibnamefont
  {Helton}}, \bibinfo {author} {\bibfnamefont {K.}~\bibnamefont {Matan}},
  \bibinfo {author} {\bibfnamefont {M.~P.}\ \bibnamefont {Shores}}, \bibinfo
  {author} {\bibfnamefont {E.~A.}\ \bibnamefont {Nytko}}, \bibinfo {author}
  {\bibfnamefont {B.~M.}\ \bibnamefont {Bartlett}}, \bibinfo {author}
  {\bibfnamefont {Y.}~\bibnamefont {Yoshida}}, \bibinfo {author} {\bibfnamefont
  {Y.}~\bibnamefont {Takano}}, \bibinfo {author} {\bibfnamefont
  {A.}~\bibnamefont {Suslov}}, \bibinfo {author} {\bibfnamefont
  {Y.}~\bibnamefont {Qiu}}, \bibinfo {author} {\bibfnamefont {J.-H.}\
  \bibnamefont {Chung}}, \emph {et~al.},\ }\href
  {https://link.aps.org/doi/10.1103/PhysRevLett.98.107204} {\bibfield
  {journal} {\bibinfo  {journal} {Phys. Rev. Lett.}\ }\textbf {\bibinfo
  {volume} {98}},\ \bibinfo {pages} {107204} (\bibinfo {year}
  {2007})}\BibitemShut {NoStop}%
\bibitem [{\citenamefont {Ghimire}\ \emph {et~al.}(2020)\citenamefont
  {Ghimire}, \citenamefont {Dally}, \citenamefont {Poudel}, \citenamefont
  {Jones}, \citenamefont {Michel}, \citenamefont {Magar}, \citenamefont
  {Bleuel}, \citenamefont {McGuire}, \citenamefont {Jiang}, \citenamefont
  {Mitchell} \emph {et~al.}}]{sciadv.abe2680}%
  \BibitemOpen
  \bibfield  {author} {\bibinfo {author} {\bibfnamefont {N.~J.}\ \bibnamefont
  {Ghimire}}, \bibinfo {author} {\bibfnamefont {R.~L.}\ \bibnamefont {Dally}},
  \bibinfo {author} {\bibfnamefont {L.}~\bibnamefont {Poudel}}, \bibinfo
  {author} {\bibfnamefont {D.~C.}\ \bibnamefont {Jones}}, \bibinfo {author}
  {\bibfnamefont {D.}~\bibnamefont {Michel}}, \bibinfo {author} {\bibfnamefont
  {N.~T.}\ \bibnamefont {Magar}}, \bibinfo {author} {\bibfnamefont
  {M.}~\bibnamefont {Bleuel}}, \bibinfo {author} {\bibfnamefont {M.~A.}\
  \bibnamefont {McGuire}}, \bibinfo {author} {\bibfnamefont {J.~S.}\
  \bibnamefont {Jiang}}, \bibinfo {author} {\bibfnamefont {J.~F.}\ \bibnamefont
  {Mitchell}}, \emph {et~al.},\ }\href
  {https://www.science.org/doi/abs/10.1126/sciadv.abe2680} {\bibfield
  {journal} {\bibinfo  {journal} {Sci. Adv.}\ }\textbf {\bibinfo {volume}
  {6}},\ \bibinfo {pages} {eabe2680} (\bibinfo {year} {2020})}\BibitemShut
  {NoStop}%
\bibitem [{\citenamefont {Wang}\ \emph {et~al.}(2021)\citenamefont {Wang},
  \citenamefont {Neubauer}, \citenamefont {Duan}, \citenamefont {Yin},
  \citenamefont {Fujitsu}, \citenamefont {Hosono}, \citenamefont {Ye},
  \citenamefont {Zhang}, \citenamefont {Chi}, \citenamefont {Krycka} \emph
  {et~al.}}]{PhysRevB.103.014416}%
  \BibitemOpen
  \bibfield  {author} {\bibinfo {author} {\bibfnamefont {Q.}~\bibnamefont
  {Wang}}, \bibinfo {author} {\bibfnamefont {K.~J.}\ \bibnamefont {Neubauer}},
  \bibinfo {author} {\bibfnamefont {C.}~\bibnamefont {Duan}}, \bibinfo {author}
  {\bibfnamefont {Q.}~\bibnamefont {Yin}}, \bibinfo {author} {\bibfnamefont
  {S.}~\bibnamefont {Fujitsu}}, \bibinfo {author} {\bibfnamefont
  {H.}~\bibnamefont {Hosono}}, \bibinfo {author} {\bibfnamefont
  {F.}~\bibnamefont {Ye}}, \bibinfo {author} {\bibfnamefont {R.}~\bibnamefont
  {Zhang}}, \bibinfo {author} {\bibfnamefont {S.}~\bibnamefont {Chi}}, \bibinfo
  {author} {\bibfnamefont {K.}~\bibnamefont {Krycka}}, \emph {et~al.},\ }\href
  {https://link.aps.org/doi/10.1103/PhysRevB.103.014416} {\bibfield  {journal}
  {\bibinfo  {journal} {Phys. Rev. B}\ }\textbf {\bibinfo {volume} {103}},\
  \bibinfo {pages} {014416} (\bibinfo {year} {2021})}\BibitemShut {NoStop}%
\bibitem [{\citenamefont {Chen}\ \emph {et~al.}(2024)\citenamefont {Chen},
  \citenamefont {Zhou}, \citenamefont {Zhang}, \citenamefont {Ji},
  \citenamefont {Liao}, \citenamefont {Ji}, \citenamefont {Li}, \citenamefont
  {Guo}, \citenamefont {Shen}, \citenamefont {Yu} \emph
  {et~al.}}]{CommunMater.024.00513}%
  \BibitemOpen
  \bibfield  {author} {\bibinfo {author} {\bibfnamefont {L.}~\bibnamefont
  {Chen}}, \bibinfo {author} {\bibfnamefont {Y.}~\bibnamefont {Zhou}}, \bibinfo
  {author} {\bibfnamefont {H.}~\bibnamefont {Zhang}}, \bibinfo {author}
  {\bibfnamefont {X.}~\bibnamefont {Ji}}, \bibinfo {author} {\bibfnamefont
  {K.}~\bibnamefont {Liao}}, \bibinfo {author} {\bibfnamefont {Y.}~\bibnamefont
  {Ji}}, \bibinfo {author} {\bibfnamefont {Y.}~\bibnamefont {Li}}, \bibinfo
  {author} {\bibfnamefont {Z.}~\bibnamefont {Guo}}, \bibinfo {author}
  {\bibfnamefont {X.}~\bibnamefont {Shen}}, \bibinfo {author} {\bibfnamefont
  {R.}~\bibnamefont {Yu}}, \emph {et~al.},\ }\href
  {https://doi.org/10.1038/s43246-024-00513-4} {\bibfield  {journal} {\bibinfo
  {journal} {Commun. Mater.}\ }\textbf {\bibinfo {volume} {5}},\ \bibinfo
  {pages} {73} (\bibinfo {year} {2024})}\BibitemShut {NoStop}%
\bibitem [{\citenamefont {Yin}\ \emph {et~al.}(2022)\citenamefont {Yin},
  \citenamefont {Lian},\ and\ \citenamefont {Hasan}}]{Nature.022.05516}%
  \BibitemOpen
  \bibfield  {author} {\bibinfo {author} {\bibfnamefont {J.-X.}\ \bibnamefont
  {Yin}}, \bibinfo {author} {\bibfnamefont {B.}~\bibnamefont {Lian}},\ and\
  \bibinfo {author} {\bibfnamefont {M.~Z.}\ \bibnamefont {Hasan}},\ }\href
  {https://doi.org/10.1038/s41586-022-05516-0} {\bibfield  {journal} {\bibinfo
  {journal} {Nature}\ }\textbf {\bibinfo {volume} {612}},\ \bibinfo {pages}
  {647} (\bibinfo {year} {2022})}\BibitemShut {NoStop}%
\bibitem [{\citenamefont {Ye}\ \emph {et~al.}(2018)\citenamefont {Ye},
  \citenamefont {Kang}, \citenamefont {Liu}, \citenamefont {von Cube},
  \citenamefont {Wicker}, \citenamefont {Suzuki}, \citenamefont {Jozwiak},
  \citenamefont {Bostwick}, \citenamefont {Rotenberg}, \citenamefont {Bell}
  \emph {et~al.}}]{nature25987}%
  \BibitemOpen
  \bibfield  {author} {\bibinfo {author} {\bibfnamefont {L.}~\bibnamefont
  {Ye}}, \bibinfo {author} {\bibfnamefont {M.}~\bibnamefont {Kang}}, \bibinfo
  {author} {\bibfnamefont {J.}~\bibnamefont {Liu}}, \bibinfo {author}
  {\bibfnamefont {F.}~\bibnamefont {von Cube}}, \bibinfo {author}
  {\bibfnamefont {C.~R.}\ \bibnamefont {Wicker}}, \bibinfo {author}
  {\bibfnamefont {T.}~\bibnamefont {Suzuki}}, \bibinfo {author} {\bibfnamefont
  {C.}~\bibnamefont {Jozwiak}}, \bibinfo {author} {\bibfnamefont
  {A.}~\bibnamefont {Bostwick}}, \bibinfo {author} {\bibfnamefont
  {E.}~\bibnamefont {Rotenberg}}, \bibinfo {author} {\bibfnamefont {D.~C.}\
  \bibnamefont {Bell}}, \emph {et~al.},\ }\href
  {https://doi.org/10.1038/nature25987} {\bibfield  {journal} {\bibinfo
  {journal} {Nature}\ }\textbf {\bibinfo {volume} {555}},\ \bibinfo {pages}
  {638} (\bibinfo {year} {2018})}\BibitemShut {NoStop}%
\bibitem [{\citenamefont {Wu}\ \emph {et~al.}(2021)\citenamefont {Wu},
  \citenamefont {Schwemmer}, \citenamefont {M\"uller}, \citenamefont
  {Consiglio}, \citenamefont {Sangiovanni}, \citenamefont {Di~Sante},
  \citenamefont {Iqbal}, \citenamefont {Hanke}, \citenamefont {Schnyder},
  \citenamefont {Denner} \emph {et~al.}}]{PhysRevLett.127.177001}%
  \BibitemOpen
  \bibfield  {author} {\bibinfo {author} {\bibfnamefont {X.}~\bibnamefont
  {Wu}}, \bibinfo {author} {\bibfnamefont {T.}~\bibnamefont {Schwemmer}},
  \bibinfo {author} {\bibfnamefont {T.}~\bibnamefont {M\"uller}}, \bibinfo
  {author} {\bibfnamefont {A.}~\bibnamefont {Consiglio}}, \bibinfo {author}
  {\bibfnamefont {G.}~\bibnamefont {Sangiovanni}}, \bibinfo {author}
  {\bibfnamefont {D.}~\bibnamefont {Di~Sante}}, \bibinfo {author}
  {\bibfnamefont {Y.}~\bibnamefont {Iqbal}}, \bibinfo {author} {\bibfnamefont
  {W.}~\bibnamefont {Hanke}}, \bibinfo {author} {\bibfnamefont {A.~P.}\
  \bibnamefont {Schnyder}}, \bibinfo {author} {\bibfnamefont {M.~M.}\
  \bibnamefont {Denner}}, \emph {et~al.},\ }\href
  {https://link.aps.org/doi/10.1103/PhysRevLett.127.177001} {\bibfield
  {journal} {\bibinfo  {journal} {Phys. Rev. Lett.}\ }\textbf {\bibinfo
  {volume} {127}},\ \bibinfo {pages} {177001} (\bibinfo {year}
  {2021})}\BibitemShut {NoStop}%
\bibitem [{\citenamefont {Tan}\ \emph {et~al.}(2021)\citenamefont {Tan},
  \citenamefont {Liu}, \citenamefont {Wang},\ and\ \citenamefont
  {Yan}}]{PhysRevLett.127.046401}%
  \BibitemOpen
  \bibfield  {author} {\bibinfo {author} {\bibfnamefont {H.}~\bibnamefont
  {Tan}}, \bibinfo {author} {\bibfnamefont {Y.}~\bibnamefont {Liu}}, \bibinfo
  {author} {\bibfnamefont {Z.}~\bibnamefont {Wang}},\ and\ \bibinfo {author}
  {\bibfnamefont {B.}~\bibnamefont {Yan}},\ }\href
  {https://link.aps.org/doi/10.1103/PhysRevLett.127.046401} {\bibfield
  {journal} {\bibinfo  {journal} {Phys. Rev. Lett.}\ }\textbf {\bibinfo
  {volume} {127}},\ \bibinfo {pages} {046401} (\bibinfo {year}
  {2021})}\BibitemShut {NoStop}%
\bibitem [{\citenamefont {Wang}\ \emph {et~al.}(2023)\citenamefont {Wang},
  \citenamefont {Wu}, \citenamefont {McCandless}, \citenamefont {Chan},\ and\
  \citenamefont {Ali}}]{NatureRevPhys.10.1038}%
  \BibitemOpen
  \bibfield  {author} {\bibinfo {author} {\bibfnamefont {Y.}~\bibnamefont
  {Wang}}, \bibinfo {author} {\bibfnamefont {H.}~\bibnamefont {Wu}}, \bibinfo
  {author} {\bibfnamefont {G.~T.}\ \bibnamefont {McCandless}}, \bibinfo
  {author} {\bibfnamefont {J.~Y.}\ \bibnamefont {Chan}},\ and\ \bibinfo
  {author} {\bibfnamefont {M.~N.}\ \bibnamefont {Ali}},\ }\href
  {https://doi.org/10.1038/s42254-023-00635-7} {\bibfield  {journal} {\bibinfo
  {journal} {Nature Rev. Phys.}\ }\textbf {\bibinfo {volume} {5}},\ \bibinfo
  {pages} {635} (\bibinfo {year} {2023})}\BibitemShut {NoStop}%
\bibitem [{\citenamefont {Higa}\ and\ \citenamefont
  {Asano}(2016)}]{PhysRevB.93.245123}%
  \BibitemOpen
  \bibfield  {author} {\bibinfo {author} {\bibfnamefont {R.}~\bibnamefont
  {Higa}}\ and\ \bibinfo {author} {\bibfnamefont {K.}~\bibnamefont {Asano}},\
  }\href {https://link.aps.org/doi/10.1103/PhysRevB.93.245123} {\bibfield
  {journal} {\bibinfo  {journal} {Phys. Rev. B}\ }\textbf {\bibinfo {volume}
  {93}},\ \bibinfo {pages} {245123} (\bibinfo {year} {2016})}\BibitemShut
  {NoStop}%
\bibitem [{\citenamefont {Seki}\ \emph {et~al.}(2018)\citenamefont {Seki},
  \citenamefont {Shirakawa},\ and\ \citenamefont
  {Yunoki}}]{PhysRevB.98.205114}%
  \BibitemOpen
  \bibfield  {author} {\bibinfo {author} {\bibfnamefont {K.}~\bibnamefont
  {Seki}}, \bibinfo {author} {\bibfnamefont {T.}~\bibnamefont {Shirakawa}},\
  and\ \bibinfo {author} {\bibfnamefont {S.}~\bibnamefont {Yunoki}},\ }\href
  {https://doi.org/10.1103/PhysRevB.98.205114} {\bibfield  {journal} {\bibinfo
  {journal} {Phys. Rev. B}\ }\textbf {\bibinfo {volume} {98}},\ \bibinfo
  {pages} {205114} (\bibinfo {year} {2018})}\BibitemShut {NoStop}%
\bibitem [{\citenamefont {Ohashi}\ \emph {et~al.}(2006)\citenamefont {Ohashi},
  \citenamefont {Kawakami},\ and\ \citenamefont
  {Tsunetsugu}}]{PhysRevLett.97.066401}%
  \BibitemOpen
  \bibfield  {author} {\bibinfo {author} {\bibfnamefont {T.}~\bibnamefont
  {Ohashi}}, \bibinfo {author} {\bibfnamefont {N.}~\bibnamefont {Kawakami}},\
  and\ \bibinfo {author} {\bibfnamefont {H.}~\bibnamefont {Tsunetsugu}},\
  }\href {https://link.aps.org/doi/10.1103/PhysRevLett.97.066401} {\bibfield
  {journal} {\bibinfo  {journal} {Phys. Rev. Lett.}\ }\textbf {\bibinfo
  {volume} {97}},\ \bibinfo {pages} {066401} (\bibinfo {year}
  {2006})}\BibitemShut {NoStop}%
\bibitem [{\citenamefont {Klett}\ \emph {et~al.}(2020)\citenamefont {Klett},
  \citenamefont {Wentzell}, \citenamefont {Sch\"afer}, \citenamefont
  {Simkovic}, \citenamefont {Parcollet}, \citenamefont {Andergassen},\ and\
  \citenamefont {Hansmann}}]{PhysRevResearch.2.033476}%
  \BibitemOpen
  \bibfield  {author} {\bibinfo {author} {\bibfnamefont {M.}~\bibnamefont
  {Klett}}, \bibinfo {author} {\bibfnamefont {N.}~\bibnamefont {Wentzell}},
  \bibinfo {author} {\bibfnamefont {T.}~\bibnamefont {Sch\"afer}}, \bibinfo
  {author} {\bibfnamefont {F.}~\bibnamefont {Simkovic}}, \bibinfo {author}
  {\bibfnamefont {O.}~\bibnamefont {Parcollet}}, \bibinfo {author}
  {\bibfnamefont {S.}~\bibnamefont {Andergassen}},\ and\ \bibinfo {author}
  {\bibfnamefont {P.}~\bibnamefont {Hansmann}},\ }\href
  {https://doi.org/10.1103/PhysRevResearch.2.033476} {\bibfield  {journal}
  {\bibinfo  {journal} {Phys. Rev. Res.}\ }\textbf {\bibinfo {volume} {2}},\
  \bibinfo {pages} {033476} (\bibinfo {year} {2020})}\BibitemShut {NoStop}%
\bibitem [{\citenamefont {Kuratani}\ \emph {et~al.}(2007)\citenamefont
  {Kuratani}, \citenamefont {Koga},\ and\ \citenamefont
  {Kawakami}}]{CondensedMatter.14.145252}%
  \BibitemOpen
  \bibfield  {author} {\bibinfo {author} {\bibfnamefont {S.}~\bibnamefont
  {Kuratani}}, \bibinfo {author} {\bibfnamefont {A.}~\bibnamefont {Koga}},\
  and\ \bibinfo {author} {\bibfnamefont {N.}~\bibnamefont {Kawakami}},\ }\href
  {https://dx.doi.org/10.1088/0953-8984/19/14/145252} {\bibfield  {journal}
  {\bibinfo  {journal} {J. Phys. Cond. Matter}\ }\textbf {\bibinfo {volume}
  {19}},\ \bibinfo {pages} {145252} (\bibinfo {year} {2007})}\BibitemShut
  {NoStop}%
\bibitem [{\citenamefont {He}\ \emph {et~al.}(2024)\citenamefont {He},
  \citenamefont {Yu},\ and\ \citenamefont {Li}}]{PhysRevLett.133.096501}%
  \BibitemOpen
  \bibfield  {author} {\bibinfo {author} {\bibfnamefont {L.-W.}\ \bibnamefont
  {He}}, \bibinfo {author} {\bibfnamefont {S.-L.}\ \bibnamefont {Yu}},\ and\
  \bibinfo {author} {\bibfnamefont {J.-X.}\ \bibnamefont {Li}},\ }\href
  {https://doi.org/10.1103/PhysRevLett.133.096501} {\bibfield  {journal}
  {\bibinfo  {journal} {Phys. Rev. Lett.}\ }\textbf {\bibinfo {volume} {133}},\
  \bibinfo {pages} {096501} (\bibinfo {year} {2024})}\BibitemShut {NoStop}%
\bibitem [{\citenamefont {Medeiros-Silva}\ \emph {et~al.}(2023)\citenamefont
  {Medeiros-Silva}, \citenamefont {Costa},\ and\ \citenamefont
  {Paiva}}]{PhysRevB.107.035134}%
  \BibitemOpen
  \bibfield  {author} {\bibinfo {author} {\bibfnamefont {A.~R.}\ \bibnamefont
  {Medeiros-Silva}}, \bibinfo {author} {\bibfnamefont {N.~C.}\ \bibnamefont
  {Costa}},\ and\ \bibinfo {author} {\bibfnamefont {T.}~\bibnamefont {Paiva}},\
  }\href {https://link.aps.org/doi/10.1103/PhysRevB.107.035134} {\bibfield
  {journal} {\bibinfo  {journal} {Phys. Rev. B}\ }\textbf {\bibinfo {volume}
  {107}},\ \bibinfo {pages} {035134} (\bibinfo {year} {2023})}\BibitemShut
  {NoStop}%
\bibitem [{\citenamefont {Ding}\ \emph {et~al.}(2025)\citenamefont {Ding},
  \citenamefont {Li},\ and\ \citenamefont {Wang}}]{CommunPhys.10.1038}%
  \BibitemOpen
  \bibfield  {author} {\bibinfo {author} {\bibfnamefont {S.}~\bibnamefont
  {Ding}}, \bibinfo {author} {\bibfnamefont {S.}~\bibnamefont {Li}},\ and\
  \bibinfo {author} {\bibfnamefont {Y.}~\bibnamefont {Wang}},\ }\href
  {https://doi.org/10.1038/s42005-025-01963-z} {\bibfield  {journal} {\bibinfo
  {journal} {Commun. Phys.}\ }\textbf {\bibinfo {volume} {8}},\ \bibinfo
  {pages} {48} (\bibinfo {year} {2025})}\BibitemShut {NoStop}%
\bibitem [{\citenamefont {Lima}\ \emph {et~al.}(2023)\citenamefont {Lima},
  \citenamefont {Medeiros-Silva}, \citenamefont {dos Santos}, \citenamefont
  {Paiva},\ and\ \citenamefont {Costa}}]{PhysRevB.108.235163}%
  \BibitemOpen
  \bibfield  {author} {\bibinfo {author} {\bibfnamefont {L.~O.}\ \bibnamefont
  {Lima}}, \bibinfo {author} {\bibfnamefont {A.~R.}\ \bibnamefont
  {Medeiros-Silva}}, \bibinfo {author} {\bibfnamefont {R.~R.}\ \bibnamefont
  {dos Santos}}, \bibinfo {author} {\bibfnamefont {T.}~\bibnamefont {Paiva}},\
  and\ \bibinfo {author} {\bibfnamefont {N.~C.}\ \bibnamefont {Costa}},\ }\href
  {https://link.aps.org/doi/10.1103/PhysRevB.108.235163} {\bibfield  {journal}
  {\bibinfo  {journal} {Phys. Rev. B}\ }\textbf {\bibinfo {volume} {108}},\
  \bibinfo {pages} {235163} (\bibinfo {year} {2023})}\BibitemShut {NoStop}%
\bibitem [{\citenamefont {Ma}\ \emph {et~al.}(2018)\citenamefont {Ma},
  \citenamefont {Zhang}, \citenamefont {Chang}, \citenamefont {Hung},\ and\
  \citenamefont {Scalettar}}]{PhysRevLett.120.116601}%
  \BibitemOpen
  \bibfield  {author} {\bibinfo {author} {\bibfnamefont {T.}~\bibnamefont
  {Ma}}, \bibinfo {author} {\bibfnamefont {L.}~\bibnamefont {Zhang}}, \bibinfo
  {author} {\bibfnamefont {C.-C.}\ \bibnamefont {Chang}}, \bibinfo {author}
  {\bibfnamefont {H.-H.}\ \bibnamefont {Hung}},\ and\ \bibinfo {author}
  {\bibfnamefont {R.~T.}\ \bibnamefont {Scalettar}},\ }\href
  {https://link.aps.org/doi/10.1103/PhysRevLett.120.116601} {\bibfield
  {journal} {\bibinfo  {journal} {Phys. Rev. Lett.}\ }\textbf {\bibinfo
  {volume} {120}},\ \bibinfo {pages} {116601} (\bibinfo {year}
  {2018})}\BibitemShut {NoStop}%
\bibitem [{\citenamefont {Li}\ \emph {et~al.}(2009)\citenamefont {Li},
  \citenamefont {Chu}, \citenamefont {Jain},\ and\ \citenamefont
  {Shen}}]{PhysRevLett.102.136806}%
  \BibitemOpen
  \bibfield  {author} {\bibinfo {author} {\bibfnamefont {J.}~\bibnamefont
  {Li}}, \bibinfo {author} {\bibfnamefont {R.-L.}\ \bibnamefont {Chu}},
  \bibinfo {author} {\bibfnamefont {J.~K.}\ \bibnamefont {Jain}},\ and\
  \bibinfo {author} {\bibfnamefont {S.-Q.}\ \bibnamefont {Shen}},\ }\href
  {https://link.aps.org/doi/10.1103/PhysRevLett.102.136806} {\bibfield
  {journal} {\bibinfo  {journal} {Phys. Rev. Lett.}\ }\textbf {\bibinfo
  {volume} {102}},\ \bibinfo {pages} {136806} (\bibinfo {year}
  {2009})}\BibitemShut {NoStop}%
\bibitem [{\citenamefont {Meng}\ \emph {et~al.}(2021)\citenamefont {Meng},
  \citenamefont {Mondaini}, \citenamefont {Ma},\ and\ \citenamefont
  {Lin}}]{PhysRevB.104.045138}%
  \BibitemOpen
  \bibfield  {author} {\bibinfo {author} {\bibfnamefont {J.}~\bibnamefont
  {Meng}}, \bibinfo {author} {\bibfnamefont {R.}~\bibnamefont {Mondaini}},
  \bibinfo {author} {\bibfnamefont {T.}~\bibnamefont {Ma}},\ and\ \bibinfo
  {author} {\bibfnamefont {H.-Q.}\ \bibnamefont {Lin}},\ }\href
  {https://link.aps.org/doi/10.1103/PhysRevB.104.045138} {\bibfield  {journal}
  {\bibinfo  {journal} {Phys. Rev. B}\ }\textbf {\bibinfo {volume} {104}},\
  \bibinfo {pages} {045138} (\bibinfo {year} {2021})}\BibitemShut {NoStop}%
\bibitem [{\citenamefont {Kang}\ \emph {et~al.}(2022)\citenamefont {Kang},
  \citenamefont {Fang}, \citenamefont {Kim}, \citenamefont {Ortiz},
  \citenamefont {Ryu}, \citenamefont {Kim}, \citenamefont {Yoo}, \citenamefont
  {Sangiovanni}, \citenamefont {Di~Sante}, \citenamefont {Park} \emph
  {et~al.}}]{Nature.021.01451}%
  \BibitemOpen
  \bibfield  {author} {\bibinfo {author} {\bibfnamefont {M.}~\bibnamefont
  {Kang}}, \bibinfo {author} {\bibfnamefont {S.}~\bibnamefont {Fang}}, \bibinfo
  {author} {\bibfnamefont {J.-K.}\ \bibnamefont {Kim}}, \bibinfo {author}
  {\bibfnamefont {B.~R.}\ \bibnamefont {Ortiz}}, \bibinfo {author}
  {\bibfnamefont {S.~H.}\ \bibnamefont {Ryu}}, \bibinfo {author} {\bibfnamefont
  {J.}~\bibnamefont {Kim}}, \bibinfo {author} {\bibfnamefont {J.}~\bibnamefont
  {Yoo}}, \bibinfo {author} {\bibfnamefont {G.}~\bibnamefont {Sangiovanni}},
  \bibinfo {author} {\bibfnamefont {D.}~\bibnamefont {Di~Sante}}, \bibinfo
  {author} {\bibfnamefont {B.-G.}\ \bibnamefont {Park}}, \emph {et~al.},\
  }\href {https://doi.org/10.1038/s41567-021-01451-5} {\bibfield  {journal}
  {\bibinfo  {journal} {Nat. Phys.}\ }\textbf {\bibinfo {volume} {18}},\
  \bibinfo {pages} {301} (\bibinfo {year} {2022})}\BibitemShut {NoStop}%
\bibitem [{\citenamefont {Blankenbecler}\ \emph {et~al.}(1981)\citenamefont
  {Blankenbecler}, \citenamefont {Scalapino},\ and\ \citenamefont
  {Sugar}}]{PhysRevD.24.2278}%
  \BibitemOpen
  \bibfield  {author} {\bibinfo {author} {\bibfnamefont {R.}~\bibnamefont
  {Blankenbecler}}, \bibinfo {author} {\bibfnamefont {D.~J.}\ \bibnamefont
  {Scalapino}},\ and\ \bibinfo {author} {\bibfnamefont {R.~L.}\ \bibnamefont
  {Sugar}},\ }\href {https://doi.org/10.1103/PhysRevD.24.2278} {\bibfield
  {journal} {\bibinfo  {journal} {Phys. Rev. D}\ }\textbf {\bibinfo {volume}
  {24}},\ \bibinfo {pages} {2278} (\bibinfo {year} {1981})}\BibitemShut
  {NoStop}%
\bibitem [{\citenamefont {White}\ \emph {et~al.}(1989)\citenamefont {White},
  \citenamefont {Scalapino}, \citenamefont {Sugar}, \citenamefont {Loh},
  \citenamefont {Gubernatis},\ and\ \citenamefont
  {Scalettar}}]{PhysRevB.40.506}%
  \BibitemOpen
  \bibfield  {author} {\bibinfo {author} {\bibfnamefont {S.~R.}\ \bibnamefont
  {White}}, \bibinfo {author} {\bibfnamefont {D.~J.}\ \bibnamefont
  {Scalapino}}, \bibinfo {author} {\bibfnamefont {R.~L.}\ \bibnamefont
  {Sugar}}, \bibinfo {author} {\bibfnamefont {E.~Y.}\ \bibnamefont {Loh}},
  \bibinfo {author} {\bibfnamefont {J.~E.}\ \bibnamefont {Gubernatis}},\ and\
  \bibinfo {author} {\bibfnamefont {R.~T.}\ \bibnamefont {Scalettar}},\ }\href
  {https://doi.org/10.1103/PhysRevB.40.506} {\bibfield  {journal} {\bibinfo
  {journal} {Phys. Rev. B}\ }\textbf {\bibinfo {volume} {40}},\ \bibinfo
  {pages} {506} (\bibinfo {year} {1989})}\BibitemShut {NoStop}%
\bibitem [{\citenamefont {Trivedi}\ \emph {et~al.}(1996)\citenamefont
  {Trivedi}, \citenamefont {Scalettar},\ and\ \citenamefont
  {Randeria}}]{PhysRevB.54.R3756}%
  \BibitemOpen
  \bibfield  {author} {\bibinfo {author} {\bibfnamefont {N.}~\bibnamefont
  {Trivedi}}, \bibinfo {author} {\bibfnamefont {R.~T.}\ \bibnamefont
  {Scalettar}},\ and\ \bibinfo {author} {\bibfnamefont {M.}~\bibnamefont
  {Randeria}},\ }\href {https://link.aps.org/doi/10.1103/PhysRevB.54.R3756}
  {\bibfield  {journal} {\bibinfo  {journal} {Phys. Rev. B}\ }\textbf {\bibinfo
  {volume} {54}},\ \bibinfo {pages} {R3756} (\bibinfo {year}
  {1996})}\BibitemShut {NoStop}%
\bibitem [{\citenamefont {Tian}\ \emph {et~al.}(2022)\citenamefont {Tian},
  \citenamefont {Li}, \citenamefont {Liang},\ and\ \citenamefont
  {Ma}}]{PhysRevB.105.045132}%
  \BibitemOpen
  \bibfield  {author} {\bibinfo {author} {\bibfnamefont {L.}~\bibnamefont
  {Tian}}, \bibinfo {author} {\bibfnamefont {Y.}~\bibnamefont {Li}}, \bibinfo
  {author} {\bibfnamefont {Y.}~\bibnamefont {Liang}},\ and\ \bibinfo {author}
  {\bibfnamefont {T.}~\bibnamefont {Ma}},\ }\href
  {https://link.aps.org/doi/10.1103/PhysRevB.105.045132} {\bibfield  {journal}
  {\bibinfo  {journal} {Phys. Rev. B}\ }\textbf {\bibinfo {volume} {105}},\
  \bibinfo {pages} {045132} (\bibinfo {year} {2022})}\BibitemShut {NoStop}%
\bibitem [{\citenamefont {Guo}\ \emph {et~al.}(2024)\citenamefont {Guo},
  \citenamefont {Liang},\ and\ \citenamefont {Ma}}]{PhysRevB.109.045107}%
  \BibitemOpen
  \bibfield  {author} {\bibinfo {author} {\bibfnamefont {K.}~\bibnamefont
  {Guo}}, \bibinfo {author} {\bibfnamefont {Y.}~\bibnamefont {Liang}},\ and\
  \bibinfo {author} {\bibfnamefont {T.}~\bibnamefont {Ma}},\ }\href
  {https://doi.org/10.1103/PhysRevB.109.045107} {\bibfield  {journal} {\bibinfo
   {journal} {Phys. Rev. B}\ }\textbf {\bibinfo {volume} {109}},\ \bibinfo
  {pages} {045107} (\bibinfo {year} {2024})}\BibitemShut {NoStop}%
\bibitem [{\citenamefont {Wang}\ \emph {et~al.}(2020)\citenamefont {Wang},
  \citenamefont {Zhang}, \citenamefont {Ma}, \citenamefont {Chen},
  \citenamefont {Liang},\ and\ \citenamefont {Ma}}]{PhysRevB.101.245161}%
  \BibitemOpen
  \bibfield  {author} {\bibinfo {author} {\bibfnamefont {J.}~\bibnamefont
  {Wang}}, \bibinfo {author} {\bibfnamefont {L.}~\bibnamefont {Zhang}},
  \bibinfo {author} {\bibfnamefont {R.}~\bibnamefont {Ma}}, \bibinfo {author}
  {\bibfnamefont {Q.}~\bibnamefont {Chen}}, \bibinfo {author} {\bibfnamefont
  {Y.}~\bibnamefont {Liang}},\ and\ \bibinfo {author} {\bibfnamefont
  {T.}~\bibnamefont {Ma}},\ }\href
  {https://doi.org/10.1103/PhysRevB.101.245161} {\bibfield  {journal} {\bibinfo
   {journal} {Phys. Rev. B}\ }\textbf {\bibinfo {volume} {101}},\ \bibinfo
  {pages} {245161} (\bibinfo {year} {2020})}\BibitemShut {NoStop}%
\bibitem [{\citenamefont {Trivedi}\ and\ \citenamefont
  {Randeria}(1995)}]{PhysRevLett.75.312}%
  \BibitemOpen
  \bibfield  {author} {\bibinfo {author} {\bibfnamefont {N.}~\bibnamefont
  {Trivedi}}\ and\ \bibinfo {author} {\bibfnamefont {M.}~\bibnamefont
  {Randeria}},\ }\href {https://link.aps.org/doi/10.1103/PhysRevLett.75.312}
  {\bibfield  {journal} {\bibinfo  {journal} {Phys. Rev. Lett.}\ }\textbf
  {\bibinfo {volume} {75}},\ \bibinfo {pages} {312} (\bibinfo {year}
  {1995})}\BibitemShut {NoStop}%
\bibitem [{\citenamefont {Iglovikov}\ \emph {et~al.}(2014)\citenamefont
  {Iglovikov}, \citenamefont {H\'ebert}, \citenamefont {Gr\'emaud},
  \citenamefont {Batrouni},\ and\ \citenamefont
  {Scalettar}}]{PhysRevB.90.094506}%
  \BibitemOpen
  \bibfield  {author} {\bibinfo {author} {\bibfnamefont {V.~I.}\ \bibnamefont
  {Iglovikov}}, \bibinfo {author} {\bibfnamefont {F.}~\bibnamefont {H\'ebert}},
  \bibinfo {author} {\bibfnamefont {B.}~\bibnamefont {Gr\'emaud}}, \bibinfo
  {author} {\bibfnamefont {G.~G.}\ \bibnamefont {Batrouni}},\ and\ \bibinfo
  {author} {\bibfnamefont {R.~T.}\ \bibnamefont {Scalettar}},\ }\href
  {https://doi.org/10.1103/PhysRevB.90.094506} {\bibfield  {journal} {\bibinfo
  {journal} {Phys. Rev. B}\ }\textbf {\bibinfo {volume} {90}},\ \bibinfo
  {pages} {094506} (\bibinfo {year} {2014})}\BibitemShut {NoStop}%
\bibitem [{\citenamefont {Pan}\ \emph {et~al.}(2023)\citenamefont {Pan},
  \citenamefont {Ma},\ and\ \citenamefont {Ma}}]{PhysRevB.107.245126}%
  \BibitemOpen
  \bibfield  {author} {\bibinfo {author} {\bibfnamefont {Y.}~\bibnamefont
  {Pan}}, \bibinfo {author} {\bibfnamefont {R.}~\bibnamefont {Ma}},\ and\
  \bibinfo {author} {\bibfnamefont {T.}~\bibnamefont {Ma}},\ }\href
  {https://doi.org/10.1103/PhysRevB.107.245126} {\bibfield  {journal} {\bibinfo
   {journal} {Phys. Rev. B}\ }\textbf {\bibinfo {volume} {107}},\ \bibinfo
  {pages} {245126} (\bibinfo {year} {2023})}\BibitemShut {NoStop}%
\bibitem [{\citenamefont {Loh}\ \emph {et~al.}(1990)\citenamefont {Loh},
  \citenamefont {Gubernatis}, \citenamefont {Scalettar}, \citenamefont {White},
  \citenamefont {Scalapino},\ and\ \citenamefont {Sugar}}]{PhysRevB.41.9301}%
  \BibitemOpen
  \bibfield  {author} {\bibinfo {author} {\bibfnamefont {E.~Y.}\ \bibnamefont
  {Loh}}, \bibinfo {author} {\bibfnamefont {J.~E.}\ \bibnamefont {Gubernatis}},
  \bibinfo {author} {\bibfnamefont {R.~T.}\ \bibnamefont {Scalettar}}, \bibinfo
  {author} {\bibfnamefont {S.~R.}\ \bibnamefont {White}}, \bibinfo {author}
  {\bibfnamefont {D.~J.}\ \bibnamefont {Scalapino}},\ and\ \bibinfo {author}
  {\bibfnamefont {R.~L.}\ \bibnamefont {Sugar}},\ }\href
  {https://link.aps.org/doi/10.1103/PhysRevB.41.9301} {\bibfield  {journal}
  {\bibinfo  {journal} {Phys. Rev. B}\ }\textbf {\bibinfo {volume} {41}},\
  \bibinfo {pages} {9301} (\bibinfo {year} {1990})}\BibitemShut {NoStop}%
\bibitem [{\citenamefont {Troyer}\ and\ \citenamefont
  {Wiese}(2005)}]{PhysRevLett.94.170201}%
  \BibitemOpen
  \bibfield  {author} {\bibinfo {author} {\bibfnamefont {M.}~\bibnamefont
  {Troyer}}\ and\ \bibinfo {author} {\bibfnamefont {U.-J.}\ \bibnamefont
  {Wiese}},\ }\href {https://link.aps.org/doi/10.1103/PhysRevLett.94.170201}
  {\bibfield  {journal} {\bibinfo  {journal} {Phys. Rev. Lett.}\ }\textbf
  {\bibinfo {volume} {94}},\ \bibinfo {pages} {170201} (\bibinfo {year}
  {2005})}\BibitemShut {NoStop}%
\bibitem [{\citenamefont {Yang}\ \emph {et~al.}(2016)\citenamefont {Yang},
  \citenamefont {Xu}, \citenamefont {Zhang}, \citenamefont {Ma},\ and\
  \citenamefont {Wu}}]{PhysRevB.94.075106}%
  \BibitemOpen
  \bibfield  {author} {\bibinfo {author} {\bibfnamefont {G.}~\bibnamefont
  {Yang}}, \bibinfo {author} {\bibfnamefont {S.}~\bibnamefont {Xu}}, \bibinfo
  {author} {\bibfnamefont {W.}~\bibnamefont {Zhang}}, \bibinfo {author}
  {\bibfnamefont {T.}~\bibnamefont {Ma}},\ and\ \bibinfo {author}
  {\bibfnamefont {C.}~\bibnamefont {Wu}},\ }\href
  {https://doi.org/10.1103/PhysRevB.94.075106} {\bibfield  {journal} {\bibinfo
  {journal} {Phys. Rev. B}\ }\textbf {\bibinfo {volume} {94}},\ \bibinfo
  {pages} {075106} (\bibinfo {year} {2016})}\BibitemShut {NoStop}%
\end{thebibliography}%

\end{document}